\documentclass[12pt]{article}
\usepackage[dvips]{graphics}
\usepackage{amsmath,amssymb,mathrsfs}

\newlength{\dinwidth}
\newlength{\dinmargin}
\begin{document}

\renewcommand{\square}{\vrule height 1.5ex width 1.2ex depth -.1ex }


\newcommand{\II}{\leavevmode\hbox{\rm{\small1\kern-3.8pt\normalsize1}}}

\newcommand{\CC}{{\mathbb C}}
\newcommand{\RR}{{\mathbb R}}
\newcommand{\NN}{{\mathbb N}}
\newcommand{\QQ}{{\mathbb Q}}
\newcommand{\TT}{{\mathbb T}}
\newcommand{\ZZ}{{\mathbb Z}}

\newcommand{\erf}{\mathop{\rm erf}}
\newcommand{\eps}{\varepsilon}
\newcommand{\esup}{\mathop{\rm ess\,sup}}


\newcommand{\CoinfM}{C_0^\infty(M)}
\newcommand{\CoinfN}{C_0^\infty(N)}
\newcommand{\Coinfd}{C_0^\infty(\RR^d\backslash\{ 0\})}
\newcommand{\Coinf}[1]{C_0^\infty(\RR^{#1}\backslash\{ 0\})}
\newcommand{\CoinX}[1]{C_0^\infty({#1})}
\newcommand{\Coin}{C_0^\infty(0,\infty)}


\newtheorem{Thm}{Theorem}[section]
\newtheorem{Def}[Thm]{Definition}
\newtheorem{Lem}[Thm]{Lemma}
\newtheorem{Prop}[Thm]{Proposition}
\newtheorem{Cor}[Thm]{Corollary}

\numberwithin{equation}{section}



\newcommand{\DD}{{\mathscr D}}
\newcommand{\EE}{{\mathscr E}}
\newcommand{\HH}{{\mathscr H}}
\newcommand{\KK}{{\mathscr K}}
\newcommand{\FF}{{\mathscr F}}

\newcommand{\DDco}{{\mathscr D}_{\rm cosp}}
\newcommand{\Aa}{{\cal A}}
\newcommand{\Ff}{{\cal F}}
\newcommand{\cU}{{\cal U}}
\newcommand{\Xx}{{\cal X}}
\newcommand{\Zz}{{\cal Z}}
\newcommand{\Ft}{{\widetilde{\Ff}}}

\newcommand{\gb}{{\boldsymbol{g}}}
\newcommand{\hb}{{\boldsymbol{h}}}
\newcommand{\jb}{{\boldsymbol{j}}}
\newcommand{\kb}{{\boldsymbol{k}}}
\newcommand{\nb}{{\boldsymbol{n}}}
\newcommand{\xb}{{\boldsymbol{x}}}
\newcommand{\Fb}{{\boldsymbol{F}}}
\newcommand{\xbo}{{\boldsymbol{x_0}}}
\newcommand{\etb}{{\boldsymbol{\eta}}}

\newcommand{\vf}{{\sf f}}
\newcommand{\vg}{{\sf g}}


\newcommand{\UU}{{\sf U}}
\newcommand{\PU}{P\cU}
\newcommand{\Ghat}{\widehat{G}}

\newcommand{\Wf}{{\mathfrak W}}
\newcommand{\Af}{{\mathfrak{A}}}

\newcommand{\Dal}{\fbox{\phantom{${\scriptstyle *}$}}}

\newcommand{\Ran}{{\rm Ran}\,}
\newcommand{\supp}{{\rm supp}\,}
\newcommand{\Span}{{\rm span}\,}
\newcommand{\Tr}{{\rm Tr}\,}
\renewcommand{\Re}{{\rm Re}\,}
\renewcommand{\Im}{{\rm Im}\,}

\newcommand{\ip}[2]{{\langle #1\mid #2\rangle}}
\newcommand{\ket}[1]{{\vert #1\rangle}}
\newcommand{\bra}[1]{{\langle #1 \mid}}
\newcommand{\stack}[2]{\substack{#1 \\ #2}}

\newcommand{\ub}{\overline{u}}
\newcommand{\vb}{\overline{v}}
\newcommand{\wb}{\overline{w}}
\newcommand{\Bb}{\overline{B}}
\newcommand{\Pb}{\overline{P}}
\newcommand{\Qb}{\overline{Q}}
\newcommand{\Xb}{\overline{X}}
\newcommand{\Xib}{\overline{\Xi}}
\newcommand{\Xxb}{\overline{\Xx}}
\newcommand{\Yb}{\overline{Y}}

\newcommand{\kt}{\widetilde{k}}

\newcommand{\fhat}{widehat{f}}
\newcommand{\hhat}{\widehat{h}}

\newcommand{\WF}{{\rm WF}\,}
\newcommand{\SL}{{\rm SL}}
\newcommand{\SU}{{\rm SU}}
\newcommand{\PSL}{{\rm PSL}}
\newcommand{\PSU}{{\rm PSU}}
\newcommand{\Con}{{\rm Con}}

\newcommand{\Had}{{\rm\sf Had}\,}

\newcommand{\LL}{{\mathcal L}}

\newcommand{\omABV}{\omega_{2AB(V)}}
\newcommand{\omABVl}{\omega^{\rm loc}_{2AB(V)}}

\newcommand{\Rr}{{\mathscr R}}

\newcommand{\spec}{{\rm spec}\,}

\newcommand{\OO}{{\mathscr O}}

\newcommand{\hs}{{\rm H.S.}}

\newcommand{\Diff}{{\rm Diff}}
\newcommand{\tDiffS}{\makebox[0pt][l]{{\scalebox{1.7}[1]{$\widetilde{\phantom{Diff}}$}}}{\rm Diff}_+(S^1)}

\newcommand{\Vect}{{\rm Vect}}
\newcommand{\Mob}{\hbox{M\"ob}}
\newcommand{\Lie}{{\rm Lie}}
\newcommand{\id}{{\rm id}}

\newcommand{\tf}{\widetilde{f}}
\newcommand{\tg}{\widetilde{g}}
\newcommand{\tB}{\widetilde{B}}
\newcommand{\tH}{\widetilde{H}}
\newcommand{\tS}{\widetilde{S}}
\newcommand{\tT}{\widetilde{T}}

\newcommand{\Sch}{\mathscr{S}}

\newcommand{\rhoo}{\mathaccent"7017\rho}
\newcommand{\Ro}{\mathaccent"7017R}

\begin{flushright} ESI Preprint 1559\end{flushright}
\noindent
\begin{center}
{ \Large \bf Quantum Energy Inequalities in two-dimensional conformal
field theory}
\\[10pt]
{\large \sc Christopher J.\ Fewster${}^{1\ast}$
 {\rm and}   Stefan Hollands${}^{2,3\dagger}$}
\\[10pt]  
                 ${}^1$ Department of Mathematics,
                 University of York,\\
                 Heslington,
                 York YO10 5DD, United Kingdom\\[4pt]
                 ${}^{\ast}$ e-mail: cjf3$@$york.ac.uk
                 \\[10pt]
                 ${}^2$\, Department of Physics, UCSB,\\
		 Broida Hall,
		 Santa Barbara, CA 93106, USA
\\[10pt]
                 ${}^3$\, Institut f\"ur Theoretische Physik,
Universit\"at G\"ottingen, \\
Friedrich-Hund-Platz 1, D-37077 G\"ottingen, Germany
\\[4pt]
                 ${}^{\dagger}$ e-mail: hollands@theorie.physik.uni-goettingen.de
		 \\[18pt]
\today
\end{center}
${}$\\[18pt]
{\small {\bf Abstract. }     
Quantum energy inequalities (QEIs) are state-independent lower bounds on
weighted averages of the stress-energy tensor, and have been established
for several free quantum field models. We present rigorous QEI bounds
for a class of interacting quantum fields, namely the unitary, positive
energy conformal field theories (with stress-energy tensor) on
two-dimensional Minkowski space. The QEI bound depends on the weight
used to average the stress-energy tensor and the central charge(s) of
the theory, but not on the quantum state. We give bounds for various
situations: averaging along timelike, null and spacelike curves, as well
as over a spacetime volume. In addition, we consider boundary conformal
field theories and more general `moving mirror' models. 

Our results hold for all theories obeying a minimal set of axioms
which---as we show---are satisfied by all models built from unitary
highest-weight representations of the Virasoro algebra. In particular, 
this includes all (unitary, positive energy) minimal models and rational conformal field theories. 
Our discussion of this issue collects together (and, in places,
corrects) various results from the literature which do not appear to
have been assembled in this form elsewhere. 
}
${}$



\section{Introduction}

In classical theories of matter, the stress-energy tensor $T_{\mu\nu}$ is usually taken to
satisfy ``energy conditions'', encoding various physical assumptions.
For example, the dominant energy condition (DEC) requires that
$T^\mu_{\phantom{\mu}\nu}v^\nu$ be a future-pointing causal (timelike or null) vector whenever $v^\nu$
is [reflecting the idea that energy-momentum should be propagated at or
below the speed of light], while the weak energy condition (WEC)
requires simply that the energy density seen by any observer is
nonnegative.  It is well-known that such conditions usually fail in quantum theoretical
models of matter to the extent that, at a given spacetime point,
the expectation value of the energy density can be made arbitrarily negative
by a suitable choice of state. If such negative energy densities could in fact
be sustained over a sufficiently large region of space and time,
then all sorts of unexpected physical phenomena ranging from exotic
spacetimes to violations of the second law of thermodynamics could
occur~\cite{MorrisThorne,Alcubierre,Ford78}.

However, it has been shown that the duration and magnitude of negative
energy density that can occur is constrained, at least in models of free
fields, by so-called ``quantum inequalities''. (We will use the more
specific term ``quantum energy inequalities'' (QEIs).) Results are known
for the free scalar~\cite{FRqis,FPstat,FTi,AGWQI,Flanagan02},
Dirac~\cite{Vollick,FVdirac,FewsterMistry},
Maxwell and Proca~\cite{FRqis,Pfenning_em,FewsterPfenning} and
Rarita--Schwinger fields~\cite{YuWu} in various levels of
generality, including some quite general and rigorous results. 
These inequalities state that the weighted average of the expected
energy density along a worldline is bounded from below by a negative
constant depending only on the weighting function used in the averaging
process, but not on the quantum state. Moreover, the bounds become more stringent 
if one increases the time interval over which the averaging is performed.
These quantum energy inequalities arguably exclude, or at least severely
constrain, the above-mentioned exotic physical phenomena (see, e.g.,~\cite{FRworm,FPwarp,Roman_review}).

Unfortunately, quantum inequalities of the above character are at present
only known for free field theories, leaving open the possibility that
physically interesting, interacting field theories might display a completely
different behavior in this regard. Thus, one should also investigate quantum
inequalities for interacting quantum field theories.

In the present paper, we take a first step in this direction, by deriving a 
sharp quantum energy inequality of the above character for arbitrary unitary, two-dimensional quantum
field theories with conformal invariance and positive Hamiltonian\footnote{We are 
also assuming, of course, that the theory has a stress tensor. Not all theories with conformal
invariance necessarily admit a stress-energy tensor~\cite{BSM90,Koster_absence}.}.
Our derivation is based on the realization that Flanagan's
bound~\cite{Flan} for a massless scalar field in two dimensions is in fact an argument
in conformal field theory. Indeed, a close inspection shows that
the essential part of his argument only relies upon the transformation law 
of the stress
energy operator under diffeomorphisms, common to all two-dimensional
unitary conformal field theories with positive Hamiltonian and a
stress-energy tensor.
As a result, our general bound differs from that for a massless scalar
field in two dimensions only by a multiplicative factor of the central
charge, $c$, of the conformal field theory under consideration (and the
possibility that the left- and right-moving portions of the
stress-energy tensor might have different central charges). 
We do not assume at any point that the theory is derived
from a Lagrangian, nor do we invoke (but certainly do not exclude) 
at any point the existence of any fields other than the stress-energy tensor.
The general arguments establishing the bound are sketched in
Section~\ref{sect:2dscalinv}, following Flanagan's argument fairly
closely. 
 
Some non-trivial issues of mainly technical nature have to be dealt with in 
order to make the argument rigorous for the class of weighting functions
 that we want to consider, and to show that the bound is sharp.
These issues mainly arise from the fact that the stress-energy
tensor in two-dimensional conformal field theory has the familiar transformation
 property for those diffeomorphisms of the left- resp. right-moving 
light-ray
that can be lifted to diffeomorphisms of the unit circle $S^1$ under the 
stereographic map.
However, in order to prove our quantum inequality bound for weighting functions
of compact support (and to show that it is sharp), one formally wants to
consider diffeomorphisms outside this class. These
difficulties were overcome in~\cite{Flan} by an appeal to general covariance.
In the setting explored in this paper, a different argument is needed,
and this is elaborated in Section~\ref{sect:QEIs}.

To make this argument, we need to have sufficient control over the unitary representations
of the diffeomorphism group of $S^1$ which enter the transformation law of the stress-tensor in the 
given CFT model. We therefore begin in Sec.~3 by specifying --- in an axiomatic fashion ---
the class of models to which our derivation applies. Our axioms are fairly
minimal, and particular models will generally have extra structure.
The main content of the axioms is that the theory should be covariant
with respect to $\tDiffS$, the universal covering group of the group
of orientation preserving diffeomorphisms of the circle, and invariant
under $\widetilde{\Mob}$, the subgroup covering the M\"obius
transformations of the circle. Each independent component of the
stress-energy tensor should correspond to an independent unitary multiplier
representation of $\tDiffS$ and the stress-energy tensor itself should be
formed from the infinitesimal generators of these representations. 
As we will see (in Sect.~\ref{sect:models}), these axioms will be loose enough to embrace a wide
range of theories: in particular, they encompass all unitary rational CFTs.
Nonetheless, they are sufficient conditions for the theory to obey QEIs.
We have also collected a number of facts about $\tDiffS$ and its
representations in Sect.~\ref{sect:axiomatic}; although much of this material is
regarded as well-known, comprehensive references seem not to exist.
Thus, our presentation may be of independent interest. 

In Section~\ref{sect:virasoro}, we verify that our axioms are
satisfied by models constructed from unitary, 
highest-weight representations of the Virasoro algebra. Here, we draw on
the results of Goodman and Wallach~\cite{GW} and Toledano
Laredo~\cite{TL} which make precise the sense in which such
representations may be `exponentiated' to unitary multiplier 
representations of $\tDiffS$. As particular models may be built as direct sums of tensor products of Virasoro
representations, it is also necessary to maintain explicit control of
the multiplier appearing in our representations, and we show that this
may be defined in terms of the Bott cocycle. We have not found a full proof
of this elsewhere in the literature.

We illustrate our main result by giving several applications in
Sect.\ref{sect:applications}. In particular, we derive QEIs valid along
worldlines, or for averaging over spacetime volumes. A peculiarity of
two-dimensional conformal field theory is that QEIs also exist for averages
along spacelike or null lines, in contrast to the situation in
four-dimensional theories~\cite{FHR,FewsterRoman03}. We also show that
similar results hold for conformal field theories in the presence of
moving boundaries (often called `moving mirrors'). Finally, we discuss
the failure of QEIs for sharply cut-off averaging functions.

In conclusion, we mention that it is not clear that quantum
energy inequalities involving averaging
along worldlines will hold in generic non-conformally invariant theories in 
two dimensions, or in interacting quantum field theories in dimensions
$d>2$. Olum and Graham~\cite{OlumGraham} have investigated a model with two nonlinearly
coupled fields, one of which is in a domain wall configuration, and
argued that a static negative energy density can be created in this
fashion, which can be made large by tuning the parameters of the model. 
For these reasons, we suggest that spacetime-averaged QEIs might be a
more profitable direction for future research (as mentioned, such QEIs
hold in our present context). If one were required to scale the spatial
support of the averaging with the temporal support, then averages of
long duration would necessarily sense the large positive energy
concentrated in the domain wall, preventing the overall average from
becoming too negative. This may suggest an 
appropriate formulation for QEIs in more general circumstances.

\section{Stress-energy densities of scale-invariant theories in two dimensions}
\label{sect:2dscalinv}

Let us begin by considering a general scale-invariant theory in two-dimensional Minkowski
space. The L\"uscher--Mack theorem~\cite{LusMack,Mack,FST} asserts\footnote{The theorem
assumes that the theory obeys Wightman's axioms~\cite{StrWigh}.} that if such a
theory possesses a symmetric and conserved stress-energy tensor field
$T^{\mu\nu}$ obeying
\begin{equation}
\int T^{\mu\,0}(x^0,x^1)\,dx^1 = P^\mu\,,
\end{equation}
where $P^\mu$ are the energy-momentum operators generating spacetime
translations, then $T^{\mu\nu}$ is traceless and the independent components
$T^{00}$ and $T^{01}$ may be expressed in terms of left- and
right-moving chiral components $T_L$ and $T_R$ which each depend on only one
lightlike variable:
\begin{eqnarray}
T^{00}(x^0,x^1) &=& T_R(x^0-x^1) + T_L(x^0+x^1)  \nonumber\\
T^{01}(x^0,x^1) &=& T_R(x^0-x^1) - T_L(x^0+x^1)\,.   
\label{eq:TandTLTR}
\end{eqnarray}
These fields have scaling dimension two, i.e.,
\begin{equation}
U(\lambda) T_L(v) U(\lambda)^{-1} =\lambda^2 T_L(\lambda v)
\label{eq:scaling}
\end{equation}
(and an analogous relation for $T_R$)
where $U(\lambda)$ is the unitary implementing the scaling 
$x^\mu \mapsto \lambda x^\mu$. Moreover, $T_L$ and $T_R$ commute with each other 
and satisfy relations of the form
\begin{equation}
[T_L(v_1),T_L(v_2)] = i\left(-T_L'(v_1)\delta(v_1-v_2)+ 2T_L(v_1)\delta'(v_1-v_2)
-\frac{c_L}{24\pi}\delta'''(v_1-v_2)\II\right)
\label{eq:Tcomm}
\end{equation}
(and similarly for $T_R$) where the constants $c_L$, $c_R$ are 
the central charges of the theory and are equal under the
additional assumption of parity invariance. These commutation relations
are closely related to
those of the Virasoro algebra, a central extension of the (complexified)
Lie algebra of $\Diff_+(S^1)$, the group of orientation preserving diffeomorphisms of
the circle. 

One of the key properties of a QFT is the spectrum condition, which, in the
present context, requires that $P^0\pm P^1$ be positive operators. It is
easy to see that
\begin{eqnarray}
P_R  &:=& \frac{1}{2}\left(P^0+P^1\right) = \int T_R(u)\,du \nonumber\\
P_L  &:=& \frac{1}{2}\left(P^0-P^1\right) = \int T_L(v)\,dv
\end{eqnarray}
generate translations along null light-rays, so that
$P_R$ generates translations along a left-moving null ray and vice versa.
Positivity of
these operators does not, however, entail that the stress-energy
densities themselves are everywhere nonnegative. On the contrary: for
any $v$ there is a sequence of unit vectors $\psi_n$ (in the ``Wightman domain''
of the theory) with
\begin{equation}
\langle T_L(v)\rangle_{\psi_n} \longrightarrow -\infty \qquad \hbox{as
$n\to\infty$}
\label{eq:ptwiseub}
\end{equation}
(of course there is a similar statement for $T_R$).\footnote{Here, as
elsewhere,  $\langle A\rangle_\psi :=
\ip{\psi}{A\psi}/\ip{\psi}{\psi}$ denotes the expectation value.} It is clearly enough
to show this for $v=0$. Let $\Omega$ be the vacuum state and write $T_L(f)=\int
T_L(v)f(v)\,dv$, where $f$ is a nonnegative test function. Now 
$\langle T_L(v)\rangle_\Omega =0$ by translation- and scale-invariance
of the vacuum, while $T_L(f)\Omega\not=0$ by the Reeh--Schlieder theorem
of Wightman theory (excluding the trivial possibility that $T_L(f)=0$
for all $f$). Defining $\varphi_\lambda=\Omega-\lambda T_L(f)\Omega$
($\lambda\in \RR$), it is now evident that
$\ip{\varphi_\lambda}{T_L(f)\varphi_\lambda} = -2\lambda
\|T_L(f)\Omega\|^2 + \lambda^2\ip{\Omega}{T_L(f)^3\Omega}$ is negative
for all sufficiently small positive $\lambda$. Hence
$\langle T_L(v)\rangle_{\varphi_\lambda}$ must assume negative values for
some point $v$, and we deduce the existence of a unit vector $\psi$ with $\langle
T_L(0)\rangle_\psi<0$. Defining $\psi_n=U(n)^{-1}\psi$ and using
Eq.~\eqref{eq:scaling}, we obtain Eq.~\eqref{eq:ptwiseub}.

Thus the stress-energy density at individual spacetime points is
unbounded from below, as is the case in many other quantum field
theories.\footnote{Arguments similar to those given here apply to any
theory (in dimension $d\ge 2$) with a scaling limit of positive scaling
dimension---see~\cite{Lisbon}.} In the following sections, we will
formulate precise conditions under which averaged stress-energy
densities such as $T_L(f)$ (for nonnegative $f$) obey state-independent
lower bounds: Quantum Energy Inequalities. Our
discussion is based on an argument given by Flanagan~\cite{Flan} for the
particular case of the massless free scalar field 
(corresponding to the case $c_L=c_R=1$). We now sketch the heart of the
argument, proceeding rather formally and leaving details aside. 
This is based on the transformation property of a chiral stress-energy
density $T$ of a conformal field theory (representing $T_L$ or $T_R$)  
under reparametrisations $v\mapsto V(v)$: 
\begin{equation}
T(v)\longrightarrow V'(v)^2 T(V(v)) - \frac{c}{24\pi}\{V,v\}\II
\label{eq:reparam} 
\end{equation}
where
\begin{equation}
\{V,v\} = 
\frac{V'''(v)}{V'(v)} - \frac{3}{2}\left(\frac{V''(v)}{V'(v)}\right)^2 =
-2\sqrt{V'(v)}\frac{d^2}{dv^2}\frac{1}{\sqrt{V'(v)}}
\,
\label{eq:Schwarz_def}
\end{equation}
is the Schwarz derivative of $V$. That is, to any non-zero vector $\psi$ there is
a vector $\psi_V$ (of the same norm) such that
\begin{equation}
\langle T(v)\rangle_\psi = V'(v)^2 \langle T(V(v))\rangle_{\psi_V} - 
\frac{c}{24\pi}\{V,v\}\,.
\end{equation}
(The infinitesimal form of this transformation law is simply
Eq.~\eqref{eq:Tcomm}.)

Now suppose we are given a nonnegative test function $H$ and choose a
reparametrisation such that $V'(v)=H(v)^{-1}$. Then
$\{V,v\}=-2H(v)^{-1/2}\frac{d^2}{dv^2}H(v)^{1/2}$ and
\begin{eqnarray}
\int H(v) \langle T(v)\rangle_\psi\,dv &=& 
\int V'(v) \langle T(V(v))\rangle_{\psi_V} \,dv +
\frac{c}{12\pi}\int \sqrt{H(v)}\frac{d^2}{dv^2}\sqrt{H(v)}\,dv
\nonumber\\
&=& \int \langle T(V)\rangle_{\psi_V} \,dV - \frac{c}{12\pi} \int 
\left(\frac{d}{dv}\sqrt{H(v)}\right)^2\,dv\,,
\end{eqnarray}
assuming that the integration by parts in the last term may be
accomplished without producing any boundary terms. Since the first term
on the right-hand side is $\langle P\rangle_{\psi_V}$, which is
nonnegative, we conclude that
\begin{equation}
\int H(v) \langle T(v)\rangle_\psi\,dv \ge -\frac{c}{12\pi} \int 
\left(\frac{d}{dv}\sqrt{H(v)}\right)^2\,dv\,
\label{eq:formalQEI}
\end{equation}
for arbitrary $\psi$.
Moreover, since $\langle P\rangle_\Omega = 0$, one expects the bound to be
attained for $\psi$ such that $\psi_V=\Omega$. 

Although the above conveys the essential ideas underlying the QEI
derivation (and differs from the scalar case only inasmuch as the
central charge is not restricted to $c=1$) one must exercise greater
care to produce a satisfactory argument. There are various reasons for
this. First, the reparametrisation rule~\eqref{eq:reparam} is expected
to hold only for those reparametrisations of $\RR$ which correspond to a
diffeomorphism of the compactified light-ray, and this will not
generally be the case for the coordinate $V$ invoked above. (Indeed the
reparametrisation is not even defined for $H$ vanishing outside a
compact interval.) Second, it is clearly necessary to dilineate the
class of $\psi$ for which the bound holds: for example, the left-hand side
does not even exist for every $\psi$! Finally, one needs to ensure that
the various formal manipulations relating to $\sqrt{H(v)}$ are
valid---this technical point conceals some subtle nuances (for example,
although $\sqrt{H}$ could be replaced by a [not
necessarily nonnegative] function which squares to $H$,
it is {\em not} the case that every smooth nonnegative function is the 
square of a smooth function~\cite{Glaeser}). 

Flanagan addressed the first two points for the scalar field by an
elegant appeal to general covariance in order to compare the theory on
the full line with a theory restricted to the interior of the support of
$H$. We have chosen not to make a parallel assumption for general
conformal field theories and instead present an alternative resolution
of the problem. The upshot is that the QEI~\eqref{eq:formalQEI} holds
(for $\psi$ in a specified domain) for any nonnegative $H$ belonging to
the Schwartz class\footnote{That is, the class of functions which, together
with their derivatives, vanish more rapidly than any inverse
power at infinity.} $\Sch(\RR)$ and with the integrand on the right-hand
side regarded as vanishing at any point where $H$ vanishes. The formal
statement and rigorous proof is given in Thm.~4.1.

\section{Axiomatic framework}\label{sect:axiomatic}

 In this section, we delineate in a mathematically precise manner the 
 class of models to which our rigorous QEI derivation in Sec.~4 applies. 
 We will state the required properties of these models
 in an axiomatic fashion and demonstrate later in Sec.~5 (by drawing together
 various results in the literature) that there actually exists a wide class of 
 models with those properties.    
 
 As we remarked in the previous section, independent components of
the stress-energy are associated with independent representations of
$\Diff_+(S^1)$, the group of orientation-preserving diffeomorphisms of
the circle. It is important for the validity of our arguments establishing 
 the QEI's to have sufficient control over these representations, especially their 
 continuity properties, as well as the spectral properties of certain generators. 
 The essence of our axioms therefore consists in specifying the 
 nature of the representations of $\Diff_+(S^1)$ that are allowed to occur
 in the given conformal field theory. In order to state these properties in a precise and 
 efficient way, we will set the stage in the following subsections 
 by recalling the salient facts about the group $\Diff_+(S^1)$ and 
 its unitary representations, especially the so-called ``unitary multiplier representations''. 
 With those facts at hand, we will then state our 
 axioms for the conformal field theories considered in this paper in Subsec.~3.3. 
 
 Some of our later arguments in Sec.~5 establishing the existence of conformal field theories
 obeying our axioms will also require us to know certain properties of the 
 phases that occur in the unitary multiplier representations. Our presentation will therefore 
include a discussion and analysis of those, even though this would not, strictly speaking, be necessary
in order to present our axioms.  
 
\subsection{Preliminaries concerning $\Diff_+(S^1)$}

\subsubsection{Group structure}

Beginning with the circle itself, $S^1$ will be regarded as the unit circle $\{z\in\CC: |z|=1\}$ in the
complex plane. Under the Cayley transform $C:z\mapsto i(1-z)/(1+z)$, the circle (less
$-1$) is mapped onto $\RR$; we will refer to this as the `light-ray
picture' in what follows. The real line will also enter as the 
universal covering group of $S^1$, via the map $\theta\mapsto
\tan\frac{1}{2}\theta$. We will call this copy of $\RR$ the `unrolled
circle' to distinguish it from the light-ray picture. 

A function $f$ on $S^1$ will be said to be
differentiable if $\RR\owns \theta\mapsto f(e^{i\theta})$ is, and the derivative $f'$ 
will be given by
\begin{equation}
ie^{i\theta}f'(e^{i\theta}) = \frac{d}{d\theta}f(e^{i\theta})\,.
\end{equation}

We may now define $\Diff_+(S^1)$ to be the group (under composition) 
of all diffeomorphisms $\sigma$ of
the circle to itself which are orientation preserving, in the sense that
$\sigma(z)$ winds once positively around the origin as $z$ does. 
We will also be concerned with its universal covering group $\tDiffS$,
which may be identified with the group of diffeomorphisms 
$\rho$ of $\RR$ obeying
\begin{equation}
\rho(\theta+2\pi)=\rho(\theta)+2\pi\,,
\label{eq:twopishift}
\end{equation} 
each such map determining a $\rhoo\in\Diff_+(S^1)$ by
\begin{equation}
\rhoo(e^{i\theta}) = e^{i\rho(\theta)}\,.
\end{equation} 

As examples, let us note three particularly important one-parameter subgroups of
$\tDiffS$, which will appear in our discussion: namely
$R_\phi$ ($\phi\in\RR$) corresponding to rotations on the circle, and 
$T_s$ ($s\in\RR$) and $D_\lambda$ ($\lambda>0$) corresponding respectively to translations and dilations on the
light-ray. On the unrolled circle, the rotations are defined by 
$R_\phi(\theta)=\theta+\phi$ [so that $\Ro_\phi(z)=ze^{i\phi}$], 
while the translations and dilations are defined by 
\begin{equation}
T_s(\theta)=
2\tan^{-1}\left(s+ \tan\frac{\theta}{2}\right)\quad \hbox{for}~\theta\in(-\pi,\pi)
\end{equation}
and 
\begin{equation}
D_\lambda(\theta)=2\tan^{-1}\left(\lambda\tan\frac{\theta}{2}\right)\quad \hbox{for}~\theta\in(-\pi,\pi)
\end{equation}
and are extended to other values of $\theta$ by Eq.~\eqref{eq:twopishift} and continuity.
In each case the principal branch of arctangent should be understood. 

The rotations and translations may be combined to obtain a further
one-parameter subgroup of interest, namely the special conformal transformations 
$S_{s}=R_\pi T_s R_\pi^{-1}$ ($s\in\RR$). We also observe that the elements 
$R_{2\pi k}$ ($k\in\ZZ$) constitute the centre of
$\tDiffS$ as a consequence of Eq.~\eqref{eq:twopishift}.

Taken together, the rotations, translations and dilations generate the
universal cover $\widetilde{\Mob}$ of $\Mob$, the group of M\"obius transformations
of $S^1$. This group will be the unbroken symmetry of conformal field theory;
as we will see, these theories are only covariant (rather than
invariant) with respect to the diffeomorphisms. M\"obius 
transformations of the circle take the form
\begin{equation}
z\mapsto \frac{\alpha z + \beta}{\overline{\beta}z+\overline{\alpha}}\,,
\label{eq:mob}
\end{equation}
where $\alpha,\beta\in\CC$ with $|\alpha|^2-|\beta|^2=1$. Noting the
invariance of Eq.~\eqref{eq:mob} under simultaneous negation of $\alpha$ and
$\beta$, we see that $\Mob\cong \PSU(1,1)=\SU(1,1)/\{\II,-\II\}$. In the
light-ray picture, elements of $\Mob$ act according to 
\begin{equation}
u\mapsto \frac{au+b}{cu+d}\,,
\label{eq:flts}
\end{equation}
for real coefficients $a,b,c,d$ with $ad-bc=1$, and this provides a group
isomorphism $\Mob\cong \PSL(2,\RR)$. 

\subsubsection{Lie group structure}\label{sect:LieGroupStructure}

Let $C^\infty(\RR;\RR)$ be the space of smooth, real-valued functions on $\RR$
equipped with the topology of uniform convergence of functions and their
derivatives of all orders,\footnote{That is, $f_k\to f$ iff
$\sup_{x\in\RR}|f^{(r)}_k(x)-f^{(r)}(x)|\to 0$ for all $r\ge 0$, where $f^{(r)}$
 is the $r$'th derivative of $f$.} which makes it into a Fr\'echet
space. We use $C^\infty_{2\pi}(\RR;\RR)$ to denote the Fre\'echet
subspace of $C^\infty(\RR;\RR)$ consisting of $(2\pi)$-periodic
functions. Now $\rho\in\tDiffS$ if and only if
$\tilde{\rho}(\theta)=\rho(\theta)-\theta$ is an element of $C^\infty_{2\pi}(\RR;\RR)$
obeying $\tilde{\rho}'(\theta)>-1$ for all $\theta$. Thus $\tDiffS$ is
an open subset of an affine translate of $C^\infty_{2\pi}(\RR;\RR)$ in
$C^\infty(\RR;\RR)$ and may therefore be endowed with the structure of a
Fr\'echet manifold modelled on $C^\infty_{2\pi}(\RR;\RR)$, with
$\rho\mapsto\tilde{\rho}$ acting as a global coordinate chart. Moreover,
the group operations of composition and inversion are smooth, so
$\tDiffS$ is in fact a Fr\'echet Lie group. The same structure can be induced on
$\Diff_+(S^1)$ by the quotient map. (Cf., for example, Sec.~6
of~\cite{Milnor} and example 4.2.6 in~\cite{Hamilton}.)

The Lie algebra of these groups,
$C^\infty_{2\pi}(\RR;\RR)$, may be conveniently regarded as the
space of real vector fields on the
circle, $\Vect_\RR(S^1)$. Indeed, given any smooth
one-parameter curve $t\mapsto \rho_t\in\tDiffS$, we
obtain a vector field $X$ on $S^1$ by
\begin{equation}
(Xg)(z) = \left.\frac{d}{dt} g(\rhoo_t(z))\right|_{t=0}\qquad
(g\in C^\infty(S^1))\,,
\end{equation}
which corresponds to the tangent vector to $\rho_t$ at $t=0$. This
vector field is said to be real because it may be expressed in the form
\begin{equation}
(Xg)(e^{i\theta}) = f(e^{i\theta})\frac{d}{d\theta}g(e^{i\theta})\qquad
(g\in C^\infty(S^1))
\end{equation}
for some real-valued $f\in C^\infty(S^1)$. For our purposes, however,  
it will be more
convenient to identify $\Vect(S^1)$ and $C^\infty(S^1)$ so that 
$f\in C^\infty(S^1)$ corresponds to the vector field $\vf\in\Vect(S^1)$
with action
\begin{equation}
(\vf g)(z)= f(z)g'(z)\,.
\end{equation}
With this identification, $\vf$ is real if and only if $f$ is invariant under
the antilinear conjugation $(\Gamma f)(z) = -z^2\overline{f(z)}$. We
will denote the space of $f\in C^\infty(S^1)$ obeying $\Gamma f=f$ by
$C^\infty_\Gamma(S^1)$. As
examples, it is straightforward to check that the tangent vector to the
curve $\phi\mapsto R_\phi$ at $\phi=0$ corresponds to the function
$z\mapsto iz$, while those of $s\mapsto T_s$ and $s\mapsto S_s$ at $s=0$ correspond to 
$z\mapsto \frac{i}{2}(1+z)^2$ and $z\mapsto -\frac{i}{2}(1-z)^2$ respectively.
All three functions are invariant under $\Gamma$, as
$\overline{z}=z^{-1}$ on the circle. 

\subsubsection{The Bott cocycle}

As already remarked, the Virasoro algebras underlying CFT are central extensions of the
complexified Lie algebra of $\Diff_+(S^1)$. At the level of groups,
these extensions are described by the Bott cocycle $B:\Diff_+(S^1)\times
\Diff_+(S^1)\to\RR$ given by\footnote{This differs slightly from the
form usually given, to which it is cohomologous, but which corresponds
to the Gel'fand--Fuks (rather than Virasoro) cocycle at the level of Lie
algebras. The form given here is drawn from~\cite{Segal81} with some
typographical errors corrected.}
\begin{equation}
B(\sigma_1,\sigma_2) = -\frac{1}{48\pi}\Re \int_{S^1} \log((\sigma_1\circ\sigma_2)'(z))
\frac{d}{dz}\log(\sigma_2'(z))\,dz\,,
\end{equation}
which lifts to a cocycle $\tB(\rho_1,\rho_2)=B(\rhoo_1,\rhoo_2)$ on $\tDiffS$.
Note that the logarithms do not introduce any ambiguity into this
formula, because $\sigma'(z)$ has winding number zero about the origin
for $\sigma\in\Diff_+(S^1)$. 

Let us now collect some properties of $B$ and $\tB$. First, it is immediate
from the definition that
\begin{equation}
B(\id,\sigma)=B(\sigma,\id) = 0\,,\qquad B(\sigma,\sigma^{-1})=0\qquad(\sigma\in\Diff_+(S^1))
\label{eq:B_elem_props}
\end{equation}
and that the cocycle property
\begin{equation}
B(\sigma_1,\sigma_2)+B(\sigma_1\sigma_2,\sigma_3) = 
B(\sigma_2,\sigma_3)+B(\sigma_1,\sigma_2\sigma_3)
\label{eq:cocycle_reln}
\end{equation}
holds for all $\sigma_1,\sigma_2,\sigma_3\in\Diff_+(S^1)$ (analogous results hold also for $\tB$).

Second, $B$ vanishes on $\Mob\times\Mob$ by the Cauchy integral formula
because the integrand is holomorphic in the unit disk in that case~\cite{Segal81}.
Similarly, $\tB$ vanishes on
$\widetilde{\Mob}\times\widetilde{\Mob}$.
Third, the following first derivatives are easily computed:
\begin{equation}
D_1\tB|_{(\id,\rho)}(\vf) = -\frac{1}{48\pi}\Re
\int_{S^1} f'(\rhoo(z))\frac{\rhoo''(z)}{\rhoo'(z)}\,dz
\end{equation}
and
\begin{equation}
D_2\tB|_{(\rho,\id)}(\vf) = -\frac{1}{48\pi}\Re
\int_{S^1} \left\{\frac{\rhoo'''(z)}{\rhoo'(z)}-\left(
\frac{\rhoo''(z)}{\rhoo'(z)}\right)^2\right\}f(z)\,dz\,,
\end{equation}
from which the second derivative
\begin{equation}
D_{12}\tB|_{(\id,\id)}(\vf,\vg) = -\frac{1}{48\pi}\Re
\int_{S^1} f'(z)g''(z)\,dz= \frac{1}{2}\omega(f,g)
\end{equation}
follows easily, where 
\begin{equation}
\omega(f,g) = \frac{1}{48\pi}\int_{S^1} \left(f(z)g'''(z)-f'''(z)g(z)\right)\,dz
\label{eq:Vircocycle}
\end{equation}
is the Virasoro cocycle, i.e., the Lie algebra cocycle corresponding to
$\tB$. Note that the integral in Eq.~\eqref{eq:Vircocycle} is automatically real for $f,g\in
C^\infty_\Gamma(S^1)$.

\subsection{Unitary multiplier representations of $\protect\tDiffS$}
\label{sect:diff_reps}

Let $\HH$ be a Hilbert space, and suppose that each
$\rho\in\tDiffS$ is assigned a unitary operator $U(\rho)$ on $\HH$
so that
\begin{equation}
U(\rho)U(\rho') = e^{ic\widetilde{B}(\rho,\rho')}U(\rho\rho')
\label{eq:multiplier}
\end{equation}
holds for all $\rho,\rho'\in\tDiffS$, where
$\widetilde{B}$ is the Bott cocycle introduced above. Then the map
$\rho\mapsto U(\rho)$ will be called a {\em unitary multiplier representation}
of $\tDiffS$ with cocycle $\widetilde{B}$ and central
charge $c$. Representations of this type will form the main component of
our axioms for CFT and we now collect some of their properties. 

We begin by noting that $U$ restricts to $\widetilde{\Mob}$ as a bona fide
unitary representation because $\widetilde{B}$ vanishes on
$\widetilde{\Mob}\times\widetilde{\Mob}$. It therefore obeys $U(\id)=\II$, and, because
we also have $\widetilde{B}(\rho,\rho^{-1})=0$ for all
$\rho\in\tDiffS$, we easily obtain
$U(\rho^{-1})=U(\rho)^{-1}$ from Eq.~\eqref{eq:multiplier}.

Now assume, in addition, that the map $\rho\mapsto
U(\rho)\psi$ is continuous for each fixed $\psi\in\HH$, i.e., the
representation is {\em strongly continuous}. This assumption permits us to
obtain the infinitesimal generators of the representation, which are
interpreted as smeared stress-energy densities. In more detail:
for each $f\in C_\Gamma^\infty(S^1)$, let $\vf\in\Vect_\RR(S^1)$ be the corresponding real
vector field and define a self-adjoint operator $\Theta(f)$ by
\begin{equation}
\Theta(f)\psi = \frac{1}{i}\left.\frac{d}{ds}
U(\exp(s\vf))\psi\right|_{s=0}
\label{eq:Tfdef}
\end{equation}
on the dense domain of $\psi$ for which the derivative
exists.\footnote{The additive group of real numbers does not admit
nontrivial smooth cocycles (see, e.g., Thm.~10.38 in~\cite{Varadarajan}). Thus, 
because $s\mapsto U(\exp(s\vf))$
is a strongly continuous unitary multiplier representation of $(\RR,+)$ with a
smooth multiplier, we may write $U(\exp(s\vf))=e^{i\alpha(s)}V(s)$
where $V(s)$ is a strongly continuous one-parameter group of unitaries
and $\alpha$ is a smooth and real-valued. Stone's theorem and the
Leibniz rule then guarantee that Eq.~\eqref{eq:Tfdef} does indeed define
a self-adjoint operator with domain equal to the set of $\psi$ for which
the derivative exists.\label{fn:stone}}
We then define
$\Theta(f)$ for arbitrary $f\in C^\infty(S^1)$ by 
$\Theta(f)=\Theta(\frac{1}{2}(f+\Gamma f))+i\Theta(\frac{1}{2i}(f-\Gamma f))$ on the
appropriate intersection of domains, so that
\begin{equation}
\Theta(f)^* = \Theta(\Gamma f)
\label{eq:adjointprop}
\end{equation}
holds on $D(\Theta(f))$. A dense domain $\DD\subset\HH$ will be called a
{\em domain of $C^1$-regularity} for $\rho\mapsto U(\rho)$ if (i) it is invariant under each $U(\rho)$ and
contained in the domain of each $D(\Theta(f))$, and (ii)
the map $f\mapsto \Theta(f)\psi$ defines a vector-valued distribution on
$C^\infty(S^1)$ for each $\psi\in\DD$. We assume henceforth that such a domain is
available, and also adopt the informal notation
\begin{equation}
\Theta(f) = \int_{S^1} f(z)\Theta(z)\,dz
\end{equation}
as a convenient book-keeping device, although $\Theta(z)$ should not be
interpreted as an operator in its own right. 
To illustrate the use of this notation, let $H$ be the generator of
the 1-parameter subgroup $R_\phi$ of $\widetilde{\Mob}$. Then 
\begin{equation}
H\psi=\frac{1}{i}\left.\frac{d}{d\phi}
U(R_\phi)\psi\right|_{\phi=0}
=\frac{1}{i}\left.\frac{d}{d\phi}
U(\exp(\phi\vf))\psi\right|_{\phi=0}
\end{equation}
for any $\psi\in \DD$, where $\vf$ is the tangent vector to $\phi\mapsto R_\phi$ at
$\phi=0$. As shown above, this corresponds
to the function $f(z)=iz$, so we write
\begin{equation}
H=\int_{S^1} iz\Theta(z)\,dz\,.
\end{equation}
Similarly, the generators $P$ and $K$ of the 1-parameter subgroups
$s\mapsto T_s$ and $s\mapsto S_s$ may be written as
\begin{eqnarray}
P &=& \frac{i}{2}\int_{S^1} (1+z)^2 \Theta(z) \,dz 
\label{eq:PtT}\\
K &=& -\frac{i}{2}\int_{S^1} (1-z)^2 \Theta(z) \,dz\,,
\end{eqnarray}
so that
\begin{equation}
H=\frac{1}{2}\left(P+K\right)\,,
\label{eq:HPK}
\end{equation}
on $\DD$, using linearity of $f\mapsto\Theta(f)\psi$.

One of the key properties we will require is the transformation law of the smeared
stress-energy densities under diffeomorphisms, provided by the following
result. 
\begin{Prop} \label{prop:transform}
Assume that $\HH$ carries a strongly continuous unitary
multiplier representation of $\tDiffS$ obeying Eq.~\eqref{eq:multiplier}
for which $\DD\subset\HH$ is a domain of $C^1$-regularity. Then
$\DD$ is a core for each $\Theta(f)$ with $f=\Gamma f$. Moreover, the $\Theta(f)$
transform according to
\begin{equation}
U(\rho) \Theta(f) U(\rho)^{-1}=\Theta(f_\rho) - \frac{c}{24\pi}\int_{S^1} \{\rhoo,z\}
f(z)\,dz\,\II\,,
\label{eq:LM_smeared}
\end{equation}
on vectors in $\DD$, for arbitrary $f\in
C^\infty(S^1)$, where $f_\rho(z) = \rhoo'(\rhoo^{-1}(z))f(\rhoo^{-1}(z))$ corresponds to the
vector field $\vf_\rho={\rm Ad}(\rho)(\vf)$. Furthermore, the commutation
relations
\begin{equation}
i[\Theta(g),\Theta(f)] = \Theta(g'f-f'g) + c\omega(g,f)\II\,,
\label{eq:TVirasoro}
\end{equation}
hold for arbitrary $f,g\in C^\infty(S^1)$, on vectors $\psi\in\DD\cap
D(\Theta(f)\Theta(g))\cap D(\Theta(g)\Theta(f))$.
\end{Prop}
{\noindent\em Remark:} Eq.~\eqref{eq:LM_smeared} may also be written in
the `unsmeared form'
\begin{equation}
U(\rho) \Theta(z) U(\rho)^{-1} = \rhoo'(z)^2 \Theta(\rhoo(z)) -\frac{c}{24\pi}\{\rhoo,z\}\II\,.
\label{eq:LMrho}
\end{equation}
{\noindent\em Proof:} That $\DD$ is a core follows from Theorem VIII.11
in~\cite{RSi} and footnote~\ref{fn:stone}. To obtain the stated transformation property,
choose $f\in C^\infty_\Gamma(S^1)$ and let $\vf$ be the corresponding
vector field. Then for any $\psi\in\DD$ and $\rho\in\tDiffS$,
\begin{eqnarray}
U(\rho) \Theta(f) U(\rho)^{-1}\psi &=&\left. \frac{1}{i}\frac{d}{ds}
U(\rho)U(\exp(s\vf))U(\rho^{-1})\psi\right|_{s=0} \nonumber\\
&=& \left. \frac{1}{i}\frac{d}{ds} e^{ic\varphi(s)}U(\rho\exp(s\vf)\rho^{-1})\psi
\right|_{s=0} \nonumber\\
&=& \left. \frac{1}{i}\frac{d}{ds} e^{ic\varphi(s)}U(\exp(s\vf_\rho))\psi
\right|_{s=0} \nonumber\\
&=& \Theta(f_\rho)\psi - c\varphi'(0)\psi\,,
\label{eq:LMrho_proof}
\end{eqnarray}
where
$\varphi(s)=\widetilde{B}(\rho,\exp(s\vf))+\widetilde{B}(\rho\exp(s\vf),\rho^{-1})$.
Using the fact that $\rho\exp(s\vf)=\exp(s\vf_\rho)\rho$, the
cocycle relation Eq.~\eqref{eq:cocycle_reln}, and the elementary
properties Eq.~\eqref{eq:B_elem_props}, $\varphi$ may be rewritten in
the form
\begin{equation}
\varphi(s)= \tB(\rho,\exp(s\vf))-\tB(\exp(s\vf_\rho),\rho)\,.
\end{equation}
It is now a straightforward exercise, using the first derivatives of
$\tB$ given in the previous subsection and the definition~\eqref{eq:Schwarz_def} of the
Schwarz derivative, to show that
\begin{equation}
\varphi'(0) = - \frac{1}{24\pi}\int_{S^1} \{\rhoo,z\}
f(z)\,dz
\end{equation}
(the integral is real because $\rhoo\in\Diff_+(S^1)$ and $f\in
C^\infty_\Gamma(S^1)$). 
Substituting this in Eq.~\eqref{eq:LMrho_proof}, we have obtained
Eq.~\eqref{eq:LM_smeared} (applied to $\psi$); the extension to
$f\in C^\infty(S^1)$ is immediate by linearity.

To obtain the Virasoro relations, we now put $\rho_s=\exp s\vg$,
where the vector field $\vg$ corresponds to some $g\in
C^\infty_\Gamma(S^1)$, and choose arbitrary $\psi,\varphi\in\DD$.
We now write
\begin{equation}
\ip{-i\Theta(g)\varphi}{\Theta(f)\psi} = \left.\frac{d}{ds}
\ip{U(\rho_{s}^{-1}\varphi}{\Theta(f)\psi}\right|_{s=0}
\end{equation}
and use Eq.~\eqref{eq:LM_smeared} (applied to $U(\rho_s)\psi\in\DD$) and
the Leibniz rule, together with 
\begin{equation}
\frac{d}{ds}
\Theta(f_{\rho_s})\varphi=\Theta\left(\frac{d}{ds} f_{\rho_s}\right)\varphi
\end{equation}
to rewrite the right-hand side.
The upshot is that Eq.~\eqref{eq:TVirasoro} holds in a quadratic form
sense on $\DD$, and hence as an identity on vectors $\psi\in\DD\cap
D(\Theta(f)\Theta(g))\cap D(\Theta(g)\Theta(f))$. The extension to
general $f,g\in C^\infty(S^1)$ is by linearity, as before. $\square$ 

The following are simple applications of the above result: 
\begin{equation}
U(D_\lambda)PU(D_\lambda)^{-1} = \lambda P\,;\quad
U(D_\lambda)KU(D_\lambda)^{-1} = \lambda^{-1} K\,;\quad
K=U(R_\pi)P U(R_\pi)^{-1}
\end{equation}
(note that the Schwarz derivative of a M\"obius transformation vanishes).
In particular, we observe that $K$ and $P$ must have the same spectrum,
which (as it is nonempty, closed and dilation-invariant) 
must be one of the four possibilities $\{0\}$, $[0,\infty)$, $(-\infty,0]$ or $\RR$.
Restricting attention to the first two cases, in which $P\ge 0$, we find
that $H\ge 0$ by Eq.~\eqref{eq:HPK}, because $H$ is thereby positive on
$\DD$, on which it is essentially self-adjoint. Conversely, if $H\ge 0$, we use the
identity\footnote{The fact that $P\ge 0$ iff $H\ge 0$ is well-known,
but is usually obtained from a detailed knowledge of the unitary
representations of $\widetilde{\Mob}$. Combine, for example, the
proof of Prop.~9.2.6 in~\cite{PS86} with the representation theory given
in~\cite{Lu76,KRY,Puk}. The approach
given here is adapted from Prop.~1 of~\cite{Koster_observables} (note
that the conventions differ slightly).}
\begin{equation}
U(D_\lambda)H U(D_\lambda)^{-1} = \frac{1}{2}\left(\lambda P + \lambda^{-1} K\right)
\end{equation}
on $\DD$ to deduce that $P\ge 0$ because 
\begin{equation}
\ip{\psi}{P\psi} =
\lim_{\lambda\to\infty}
\lambda^{-1}\ip{U(D_\lambda)^{-1}\psi}{HU(D_\lambda)^{-1}\psi}\ge 0
\end{equation} 
for all $\psi\in\DD$, which is again a core for $P$. Clearly, $P=0$ if and
only if $H=0$, so $\spec(P)=[0,\infty)$ if and only if $H$ is a non-zero
positive operator.

\subsection{Axioms}\label{sect:axioms}

We now come to the statement of the axioms we shall adopt for conformal
field theory. These are to be regarded as minimal requirements: specific models
will have more structure and possibly an enlarged symmetry group. 
Nonetheless, the following axioms are already
sufficient to establish the QEIs, and are satisfied in models built from
Virasoro representations (see Sect.~\ref{sect:virasoro}). Note that, as
they include the assumptions of
Sect.~\ref{sect:diff_reps}, all the conclusions of that subsection apply
to such theories, particularly Prop.~\ref{prop:transform}.

For simplicity, we state our axioms for a conformal field theory with a
single component of stress-energy; at the end of this section we describe
the (straightforward) extension to two independent components. 

\bigskip
{\noindent\bf A.~Hilbert space, diffeomorphism group and energy
positivity}
\begin{enumerate}
\item The Hilbert space $\HH$ of the theory carries a strongly continuous unitary
multiplier representation $\rho\mapsto U(\rho)$ of $\tDiffS$ obeying
Eq.~\eqref{eq:multiplier}, with central charge $c>0$. 

\item Up to phase there is a unique
unit vector $\Omega\in\HH$ which is invariant under the restriction of
$U$ to $\widetilde{\Mob}$, and which will be called the vacuum vector. 

\item The generator $P$ of the one-parameter translation subgroup $s\mapsto U(T_s)$
is assumed to be a positive self-adjoint operator. (An equivalent
requirement is that the generator $H$ of the rotation subgroup $\phi\mapsto
U(R_\phi)$ be positive, by the remarks above.)
\end{enumerate}

{\noindent\bf B.~Stress-energy density} 

The (smeared) stress-energy density $\Theta(f)$ is defined as the generator of $U(\rho)$,  
as described in the previous subsection, see Eq.~\eqref{eq:Tfdef}. 
We assume that $\HH$ contains a dense subspace $\DD\subset\HH$ such that:
\begin{enumerate}
\item $\DD$ is invariant under each $U(\rho)$, contains $\Omega$
and is contained in each $D(\Theta(f))$ for all $f\in C^\infty(S^1)$.

\item For each $\psi\in\DD$, the map $f\mapsto \Theta(f)\psi$ is a vector-valued 
distribution on $C^\infty(S^1)$ (equipped with its usual
topology of uniform convergence of functions and all their derivatives).
Thus, $\DD$ is a domain of $C^1$-regularity in the sense introduced
above. 
\item For each $\psi\in\DD$,  $\langle \Theta(z)\rangle_\psi$ is smooth on
$S^1$. 
\end{enumerate}

Given a theory of the above type living on a circle, we may define a
stress-energy density $T(v)$ living on a light-ray by the `unsmeared' formula
\begin{equation}
T(v) = \left(\frac{dz}{dv}\right)^2 \Theta(z(v))
=-\frac{4}{(1-iv)^4}\Theta(z(v))\,,
\label{eq:T_from_Theta}
\end{equation}
where
\begin{equation}
z(v) =C^{-1}(v) = \frac{1+iv}{1-iv}
\end{equation}
maps $\RR$ to $S^1$ (less $-1$, which represents the `point at
infinity'). The class of allowed smearing functions in this picture
consists of all $F\in C^\infty(\RR)$ for which $z\mapsto
\frac{i}{2}(1+z)^2 F(C(z))$ is
smooth on $S^1$ [with an appropriate limiting definition at $z=-1$]. As before,
we use an integral notation to denote such smearings, thus, for example, 
the relationship Eq.~\eqref{eq:PtT} now reads
\begin{equation}
P = \int T(v)\,dv\,.
\label{eq:PasintT}
\end{equation}
We may also deduce from axiom B.3 and Eq.~\eqref{eq:T_from_Theta} that $\langle T(v)\rangle_\psi$ decays
as $O(v^{-4})$ as $|v|\to\infty$ for $\psi\in\DD$. 

Finally, suppose $\rho\in\tDiffS$ fixes the point at infinity,
i.e., $\rhoo(-1)=-1$, and define a reparametrisation
$v\mapsto V(v)$ of $\RR$ implicitly by $z(V(v))=\rhoo(z(v))$. Then 
the transformation law Eq.~\eqref{eq:LMrho} becomes
\begin{equation}
U(\rho) T(v)U(\rho)^{-1} = V'(v)^2 T(V(v)) - \frac{c}{24\pi}\{V,v\}\II\,.
\label{eq:LMv} 
\end{equation}
Here, we have used the chain rule for Schwarz derivatives
\begin{equation}
\{z,x\} = \{z,y\} \left( \frac{dy}{dx} \right)^2 + \{y,x\}\,,
\end{equation}
where $z = z(y), y = y(x)$, 
and the fact that the Schwarz derivative of a M\"obius transformation
vanishes identically, so $\{z(v),v\}=0$.

The above structure is already enough to
encompass an interesting class of theories in Minkowski space: namely,
boundary conformal field theories (see, e.g.,~\cite{Zuber}, or~\cite{RehrenLongo} for a
recent treatment in terms of algebraic quantum field theory). In these
theories, there is a single underlying representation $U$ of $\tDiffS$ with
corresponding stress-energy density $T$, and the theory lives on
the right-hand half $x^1>0$ of Minkowski space with stress-energy tensor
given by Eq.~\eqref{eq:TandTLTR} where $T_L=T_R=T$. In particular,
$T^{01}$ vanishes on the timelike line $x^1=0$, reflecting the boundary
condition that no energy should flow out of the half-space
$x^1>0$.

A more general class of theories corresponds to the `moving mirror'
models studied in~\cite{FullingDavies} (for particular case of the massless scalar field).
Instead of an inertial boundary $x^1=0$, we consider a moving boundary
with trajectory $v=p(u)$, where $u=x^0-x^1$ and $v=x^0+x^1$ are null
coordinates on Minkowski space. The theory is defined on the portion of
Minkowski space to the right of this curve, i.e., $v>p(u)$. 
Restricting, for simplicity,
to the case in which $u\mapsto p(u)$ lifts to an element
$\rho\in\tDiffS$, the stress-energy tensor is again defined by
Eq.~\eqref{eq:TandTLTR}, where we now put
\begin{equation}
T_L(v)= T(v),\qquad T_R(u) = U(\rho)T(u)U(\rho)^{-1}\,.
\label{eq:mirrors}
\end{equation}
(Boundary CFT corresponds, of course, to the case $p(u)=u$ and hence $U(\rho)=\II$.)
It follows Eq.~\eqref{eq:LMv} 
and $\langle T(v) \rangle_\Omega$ that the energy density in the vacuum state $\Omega$ is then
\begin{equation}
\langle T_{00}(x^0,x^1)\rangle_\Omega =-\frac{c}{24\pi}\{p,u\}=\frac{c}{12\pi}\sqrt{p'(u)}\frac{d^2}{du^2}
 \frac{1}{\sqrt{p'(u)}}\,,
\end{equation}
which reduces to the result of~\cite{FullingDavies} in the case $c=1$. 
In fact the moving mirror spacetime is conformally related to the
boundary spacetime considered above (under the transformation $(u,v)\mapsto
(p(u),v)$) and this dictates the form of Eq.~\eqref{eq:mirrors},
together with the boundary condition that $\Omega$ should be the `in' vacuum
at past null infinity. It is intended to discuss this more fully elsewhere.

Conformal field theories on the whole of Minkowski space must have two
independent components of stress-energy, by the L\"uscher--Mack theorem
(see Sec.~\ref{sect:2dscalinv}).
We now briefly explain the 
required modifications to our axioms to permit the description of this
situation. There are now two commuting projective
unitary representations $U_L$ and $U_R$ of $\tDiffS$ each
restricting to $\widetilde{\Mob}$ as a unitary representation. We assume the
existence of a unique vacuum vector $\Omega$ invariant under both copies of
$\widetilde{\Mob}$ and assume that the two translation generators $P_L$, $P_R$ 
are positive. The domain $\DD$ is assumed to be invariant under both
$U_L$ and $U_R$, and each representation is generated (in the sense of
Eq.~\eqref{eq:Tfdef}) by a
corresponding stress-energy density $\Theta_L$, $\Theta_R$, each of which
obeys the regularity assumptions of axiom B. Each stress-energy density transforms according to the
Eq.~\eqref{eq:LMrho} (with central charge $c_L$ or $c_R$
as appropriate) under the
corresponding representation of $\tDiffS$ but is invariant under
the adjoint action of the other copy. We also define light-ray fields
$T_L$ and $T_R$ in the same way as above, and then define the
stress-energy tensor by Eq.~\eqref{eq:TandTLTR}.  
In particular, one may construct such a theory as a tensor
product of two conformal field theories with a single component of
stress-energy, but this is by no means the only possibility. 
 
Clearly, we could envisage theories with any number of independent components of
stress-energy in a similar fashion, but the interpretation as a theory
in Minkowski space is no longer clear. 

\section{Quantum Energy Inequalities in CFT}
\label{sect:QEIs}

\subsection{Main result}

We are now in a position to state our main result. The notation is as in
the previous section. 
\begin{Thm} \label{thm:main}
Consider a conformal field theory with a single component $T$ of
stress-energy.  
For any nonnegative $G\in\Sch(\RR)$, the quantum energy inequality
\begin{equation}
\int G(v)\langle T(v)\rangle_\psi \,dv \ge -\frac{c}{12\pi}\int
\left( \frac{d}{dv}\sqrt{G(v)} \right)^2
\,dv
\label{eq:axiomaticQI}
\end{equation}
holds for all $\psi\in\DD$, where the derivative $d/dv \, \sqrt{G}$ is 
defined to be zero for points at which $G$ vanishes:
\begin{equation}
\label{distrder}
\frac{d}{dv} \, \sqrt{G(v)} = \left\{\begin{array}{cl} G'(v)/(2\sqrt{G(v)}) &
G(v)\not=0\\ 0 & G(v)=0\,.\end{array}\right.
\end{equation}
Moreover, this bound is sharp: the right-hand side is the infimum of the left-hand side
as $\psi$ varies in $\DD$.

In a conformal field theory with two independent components of stress-energy, both $T_L$ and $T_R$
obey bounds of the above type (with weight functions $G_L,
G_R\in\Sch(\RR)$) which are simultaneously sharp in the
sense that there is a sequence of non-zero vectors $\psi_n\in\DD$ with 
\begin{eqnarray}
\int G_L(v)\langle T_L(v)\rangle_{\psi_n} \,dv &\longrightarrow& -\frac{c_L}{12\pi}\int
\left( \frac{d}{dv}\sqrt{G_L(v)} \right)^2
\,du \nonumber\\
\int G_R(u)\langle T_R(u)\rangle_{\psi_n} \,du &\longrightarrow& -\frac{c_R}{12\pi}\int
\left( \frac{d}{du}\sqrt{G_R(u)} \right)^2
\,dv 
\label{eq:simult}
\end{eqnarray}
as $n\to\infty$. 
\end{Thm}
{\noindent\em Remarks:} 
1) It is proved in Corollary~\ref{Cor:W1} in the Appendix that the square root $\sqrt{G}$ 
of a non-negative Schwartz function is in fact a distribution in the Sobolev space $W^1(\RR)$ 
(i.e., has square-integrable first derivative) and that the above 
rule~\eqref{distrder} for defining its derivative coincides with the 
usual notion of the distributional (or ``weak'') derivative of such 
a distribution. In particular, this formally establishes that the 
integral representing our QEI bound on the right side of Eq.~\eqref{eq:axiomaticQI} 
is actually finite even for smearing functions $G$ that 
are not {\em strictly} positive.\\
2) As $\DD$ is a core for any smeared energy density the QEIs can be
stated as operator inequalities, e.g.,
\begin{equation}
\int G(v) T(v) \,dv \ge -\frac{c}{12\pi}\int
\left( \frac{d}{dv}\sqrt{G(v)} \right)^2 
\,dv\II
\end{equation}
by standard quadratic form arguments (see, e.g., Theorem X.23
in~\cite{RSii}). The fact that QEIs for $T_L$ and $T_R$ are
simultaneously sharp is simply the statement that the pair formed by the
two bounds in Eq.~\eqref{eq:simult} belongs to the joint spectrum of the two operators
concerned.\\
3) The above results can of course be transformed to give QEIs on the field
$\Theta$ on the circle; one
can also follow the general strategy given below to derive QEIs based on
positivity of $H$ (rather than $P$), which would be more natural in that
setting. In addition, the results can be extended to any number of
independent stress-energy operators. We will not pursue these directions
here.

\medskip
{\noindent\em Proof:} The proof is broken down into various stages. 
We start with the case in which the nonnegative
function $G$ is smooth and compactly supported, and then extend to the
Schwartz class. As mentioned above, the
obstruction to a straightforward use of the argument summarised in Sec.~\ref{sect:2dscalinv}
is that the equation $V'(v)=1/G(v)$ does not define a diffeomorphism
which can be lifted to the circle. To circumvent this problem, we modify
$G$ to a function $H_{\epsilon, n}$ depending upon regulators $\epsilon$ and $n$. 
The function $H_{\epsilon, n}$ is constructed in such a way that the formal 
argument given Sec.~\ref{sect:2dscalinv} holds rigorously, and so that 
the desired bound is obtained as the regulators are removed. 

The two regulators have the following effect. First, we add the constant $\epsilon$ to $G(v)$, thus
obtaining a reparametrisation of the whole line by
$V'(v)=1/(G(v)+\epsilon)$. Although this reparametrisation fixes the
point at infinity, it does not lift to a diffeomorphism of the circle as
it has a discontinuous second derivative at $z=-1$ (unless $G$ is
identically zero). The remedy is to subtract from $G(v)+\epsilon$ a small compactly
supported correction, which is translated to the
right (and slightly rescaled) as $n$ increases. As noted following
Eq.~\eqref{eq:PasintT}, $\langle T(v)\rangle_\psi = O(v^{-4})$ as $v\to\infty$
for $\psi\in\DD$, and we can exploit this decay to control the limit
$n\to\infty$. Other approaches to this issue are probably
possible.\footnote{As we were completing this paper, Carpi and Weiner
released a preprint~\cite{CarpiWeiner} in which they point out that
certain nonsmooth smearings of the stress-energy density also
yield self-adjoint operators. It is likely that one could use this to find
a unitary implementation of the
reparametrisation of the line defined by $V'(v)=1/(G(v)+\epsilon)$,
removing the need for the second stage of regulation.}

The construction and properties of $H_{\epsilon, n}$ are summarised by the following lemma,
whose proof is deferred to the end of this section.
\begin{Lem} \label{lem:xine}
Given a nonnegative $G\in\CoinX{\RR}$, let
\begin{equation}
\lambda_\epsilon = \frac{1}{|\supp G|} \int \frac{G(v)}{G(v)+\epsilon}
\,dv\,,
\end{equation}
where $|\supp G|$ denotes the Lebesgue measure of the support of $G$.
Then $\lambda_\epsilon$ increases as $\epsilon\to 0^+$, with
$\lim_{\epsilon\to 0^+}\lambda_\epsilon=1$. Let 
$\eta\in\CoinX{\RR}$ obey $0\le \eta(v)\le 1/2$ for all $v$ and 
\begin{equation}
\int \frac{\eta(v)}{1-\eta(v)}\,dv = |\supp G|\,,
\end{equation}
and set
\begin{equation}
\eta_{n,\epsilon}(v) = \eta\left(\frac{v-n}{\lambda_\epsilon}\right).
\end{equation}
Then there exists an $n_0$
such that, for all $n\ge n_0$ and $\epsilon\in(0,1)$, 
\begin{enumerate}
\item the support of
$\eta_{n,\epsilon}$ lies to the right of $\supp G$ 
\item there is a diffeomorphism $\rho_{n,\epsilon}\in\tDiffS$
corresponding to a reparametrisation $v\mapsto V_{n,\epsilon}(v)$ of
the light-ray with
\begin{equation}
V_{n, \epsilon}'(v) = \frac{1}{H_{n, \epsilon}(v)},
\label{eq:VGxine}
\end{equation}
where $H_{n, \epsilon}(v) = G(v)+\epsilon(1-\eta_{n,\epsilon}(v))$.
\end{enumerate}
\end{Lem}

Now let $\psi\in\DD$ be arbitrary, so $\langle
T(v)\rangle_\psi=O(v^{-4})$ as $v\to\infty$ for the reasons mentioned above. Then the formal calculation of
Sec.~\ref{sect:2dscalinv} holds rigorously if $H$ is replaced 
by the function $H_{n,\epsilon}$ given in item (2) of the above lemma, and if 
$\psi_V$ is replaced by $U(\rho_{n,\epsilon})\psi$. This yields 
\begin{equation}
\int H_{n,\epsilon}(v)\langle T(v)\rangle_\psi \ge -\frac{c}{12\pi}\int
\left(\frac{d}{dv}\sqrt{H_{n,\epsilon}(v)}\right)^2\,dv, 
\end{equation}
the required integration by parts being valid because $H_{n,\epsilon}$ is
constant outside a compact interval. For $n\ge n_0$, the supports of
$G$ and $\eta_{n,\epsilon}$ are disjoint by item (1) of the lemma, so the integral on the
right-hand side falls into two pieces
\begin{eqnarray}
4\int \left(\frac{d}{dv}\sqrt{H_{n,\epsilon}(v)}\right)^2\,dv &=&
\int \frac{G'(v)^2}{G(v)+\epsilon}\,dv + \epsilon
\int \frac{\eta_{n,\epsilon}'(v)^2}{1-\eta_{n,\epsilon}(v)}\,dv \nonumber\\
&=&\int \frac{G'(v)^2}{G(v)+\epsilon}\,dv + \frac{\epsilon}{\lambda_\epsilon}
\int \frac{\eta'(v)^2}{1-\eta(v)}\,dv \,.
\end{eqnarray}
On the other hand, we have
\begin{equation}
\int H_{n,\epsilon}(v)\langle T(v)\rangle_\psi = \int G(v)\langle
T(v)\rangle_\psi\,dv + \epsilon\langle P\rangle_\psi -
\epsilon \int \eta_{n,\epsilon}(v) \langle T(v)\rangle_\psi\,.
\end{equation}
As $n\to\infty$, $\eta_{n,\epsilon}$ is translated off to infinity, so
the last term drops out in the limit owing to the decay of $\langle
T(v)\rangle_\psi$ as $v\to\infty$. We therefore have
\begin{equation}
\int G(v)\langle
T(v)\rangle_\psi\,dv \ge 
-\frac{c}{48\pi}\int \frac{G'(v)^2}{G(v)+\epsilon}\,dv
-\frac{\epsilon c}{48\pi\lambda_\epsilon}\int \frac{\eta'(v)^2}{1-\eta(v)}\,dv 
-\epsilon\langle P\rangle_\psi
\,,
\end{equation}
and the limit $\epsilon\to 0^+$ yields the QEI~\eqref{eq:axiomaticQI},
owing to Corollary~\ref{Cor:W1} in the Appendix and the fact that $\psi$
was an arbitrary element of $\DD$.

\bigskip

We now turn to the case in which $G$ is a nonnegative function of Schwartz class.
According to Corollary~\ref{Cor:W1}, $\sqrt{G}$ belongs to the Sobolev space
$W^1(\RR)$. It follows that we may find nonnegative $h_k\in\CoinX{\RR}$ with
$h_k\to\sqrt{G}$ and $h_k' \to d/dv \, \sqrt{G}$ in $L^2(\RR)$ as $k\to\infty$
(the derivative $d/dv \sqrt{G}$ being understood in the sense of distributions).
Thus for each $\psi\in\DD$ and $k$, we have 
\begin{equation}
\int \langle T(v)\rangle_\psi h_k(v)^2\,dv \ge -\frac{c}{12\pi}
\int h_k'(v)^2\,dv\,.
\end{equation}
In the limit $k\to\infty$ the right-hand side clearly converges to
$-c/(12\pi) \int (d/dv\,\sqrt{G})^2 \, dv$, while the left-hand side converges to 
$\int \langle T(v)\rangle_\psi G(v)\,dv$ because 
$\langle T(v)\rangle_\psi$ is bounded in $v$. The QEI~\eqref{eq:axiomaticQI}
therefore holds for all nonnegative $G\in\Sch(\RR)$. 

\bigskip

{}To show that the bound is sharp, we employ another lemma:
\begin{Lem} \label{lem:optcpct}
If $F\in\Sch(\RR)$ and $G\in\CoinX{\RR}$ are nonnegative, then
\begin{equation}
\inf_{\psi\in\DD} \int F(v) \langle
T(v)\rangle_{\psi}\,dv \le -\frac{c}{12\pi}\int
\left(\frac{d}{dv}\frac{F(v)}{\sqrt{G(v)+\epsilon}}\right)
\left(\frac{d}{dv}\sqrt{G(v)+\epsilon}\right)\,dv\,.
\label{eq:optcpct}
\end{equation}
\end{Lem}
{\noindent\em Proof:} Using the notation of Lemma~\ref{lem:xine}, let $n>n_0$ and 
$\epsilon>0$, and define
$\psi_{n,\epsilon}=U(\rho_{n,\epsilon})^{-1}\Omega$ in terms of $G$.
Since $\langle T(V_{n,\epsilon}(v))\rangle_\Omega$ vanishes
identically, the transformation law~Eq.~\eqref{eq:LMv} gives
\begin{eqnarray}
\langle T(v)\rangle_{\psi_{n,\epsilon}} &=&
-\frac{c}{24\pi}\{V_{n,\epsilon},v\} = \frac{c}{12\pi}\,\frac{1}{\sqrt{H_{n,\epsilon}(v)}}
\frac{d^2\sqrt{H_{n,\epsilon}(v)}}{dv^2} \nonumber\\
&=&
\frac{c}{12\pi}\left(\frac{1}{\sqrt{G(v)+\epsilon}}
\frac{d^2\sqrt{G(v)+\epsilon}}{dv^2}+
\frac{1}{\sqrt{1-\eta_{n,\epsilon}(v)}}\frac{d^2\sqrt{1-\eta_{n,\epsilon}(v)}}
{dv^2}\right) \nonumber\\
&&
\end{eqnarray}
because $G$ and $\eta_{n,\epsilon}$ have
disjoint supports. Note that the effect of increasing $n$ is merely to
translate the final term to the right. This term therefore vanishes in
the limit $n\to\infty$ when we integrate against $F$, because it is
pushed off into the tail of $F$. Thus we have
\begin{equation}
\lim_{n\to\infty} \int F(v) \langle
T(v)\rangle_{\psi_{n,\epsilon}}\,dv = \frac{c}{12\pi}\int
\frac{F(v)}{\sqrt{G(v)+\epsilon}}\frac{d^2}{dv^2}\sqrt{G(v)+\epsilon}
\,dv
\end{equation}
and Eq.~\eqref{eq:optcpct} is obtained after integration by parts. $\square$

Now suppose that $G$ is a nonnegative Schwartz-class function and set
$G_n(v)=\chi(v/n)G(v)$, where $\chi\in\CoinX{\RR}$, $0\le\chi(x)\le 1$
and $\chi(x)=1$ for $|x|\le 1$. One may verify that
\begin{equation}
\lim_{m\to\infty} \frac{d}{dv}\frac{G(v)}{\sqrt{G_m(v)+\epsilon}}
= \frac{d}{dv}\sqrt{G(v)+\epsilon} = \lim_{m\to\infty} \frac{d}{dv}\sqrt{G_m(v)+\epsilon}
\label{eq:limits}
\end{equation}
in $L^2(\RR)$. Applying Lemma~\ref{lem:optcpct} with $F$ and $G$
replaced by $G$ and $G_m$ respectively, these limits and the
continuity of the right-hand side of Eq.~\eqref{eq:optcpct} in both
factors [it is effectively an $L^2$-inner product] yield
\begin{equation}
\inf_{\psi\in\DD} \int G(v) \langle
T(v)\rangle_{\psi}\,dv \le -\frac{c}{12\pi}\int
\left(\frac{d}{dv}\sqrt{G(v)+\epsilon}\right)^2\,dv\,.
\end{equation}
On taking $\epsilon\to 0^+$, we conclude that the bound Eq.~\eqref{eq:axiomaticQI}
is sharp. 

\bigskip

Turning to conformal field theories with two independent components of
stress-energy, it is immediate
from the above that both $T_L$ and $T_R$ satisfy QEIs of the form
required. That the bounds are simultaneously sharp follows from the
fact that each stress-energy density transforms under its corresponding
copy of $\tDiffS$ but is invariant under the other copy. Thus the
construction used to establish sharpness of the QEI~\eqref{eq:axiomaticQI} may
be adapted in a straightforward fashion to prove Eq.~\eqref{eq:simult}. 
This concludes the proof of our main theorem~4.1. $\square$

It remains to establish the lemma used above.

{\noindent\em Proof of Lemma~\ref{lem:xine}:} It is clear (e.g., by
monotone convergence) that $\lambda_\epsilon$ increases to unity as $\epsilon\to 0^+$. 
Thus the support of $\eta_{n,\epsilon}$ will lie to the right of $\supp
G$ for all $n$ greater than some $n_0$ and all $\epsilon\in(0,1)$. We define
\begin{equation}
V_{n,\epsilon}(v) = \int_0^v \frac{1}{H_{n,\epsilon}(v')}\,dv'\,,
\end{equation}
which evidently satisfies Eq.~\eqref{eq:VGxine} and, as it is smooth and
strictly increasing with $\lim_{v\to\pm\infty}V_{n,\epsilon}(v)=\pm\infty$ gives a
diffeomorphism of $\RR$. We wish to see that this diffeomorphism can be
extended to the circle. Suppose the support of $G$ is contained within
$[-R,R]$ for some $R>0$ and that $n>n_0$. Then, for $v<-R$ we have 
\begin{equation}
V_{n,\epsilon}(v) = \frac{v}{\epsilon} + \alpha\,,
\end{equation}
where
\begin{equation}
\alpha = \frac{R}{\epsilon}+\int_0^{-R} \frac{1}{G(v)+\epsilon}\,dv\,.
\end{equation}
Now choose $S$ to the right of $\supp \eta_{n,\epsilon}$, so $\supp
\eta_{n,\epsilon}\subset (R,S)$. Then, for $v>S$
we have
\begin{eqnarray}
V_{n,\epsilon}(v) &=& \frac{v}{\epsilon} - \frac{S}{\epsilon} + \int_0^S
\frac{1}{G(v)+\epsilon(1-\eta_{n,\epsilon}(v))}\,dv \nonumber\\
&=& \frac{v}{\epsilon} + \alpha\,,
\end{eqnarray}
which follows after a small amount of calculation using the definitions
of $\eta$ and $\lambda_\epsilon$. 

Thus $v\mapsto V_{n,\epsilon}(v)$ differs from the M\"obius transformation $v\mapsto
v/\epsilon+\alpha$ only on a compact set and may therefore be lifted to
$\rho_{n,\epsilon}\in\tDiffS$ defined by 
$\rho_{n,\epsilon}(\theta)=2\tan^{-1}\left(V_{n,\epsilon}(\tan\frac{1}{2}\theta)\right)$
for $\theta\in(-\pi,\pi)$ and extended to other values by continuity and
Eq.~\eqref{eq:twopishift}. $\square$

\subsection{Applications}\label{sect:applications}

We now use Theorem~\ref{thm:main} to give various useful QEI bounds
for conformal field theories (on two-dimensional Minkowski space).

\subsubsection{Worldline bounds}\label{sect:worldlines}

Consider a smooth curve $\lambda \to \gamma^\mu(\lambda)$ in Minkowski
space, and set $u = \gamma^0 - \gamma^1$, $v = \gamma^0 + \gamma^1$.
It is straightforward to show that 
\begin{equation}
\rho_\gamma(\lambda):= 
T_{\mu\nu}(\gamma(\lambda))\dot{\gamma}^\mu(\lambda)\dot{\gamma}^\nu(\lambda)  
= T_R(u(\lambda))\dot{u}(\lambda)^2+ T_L(v(\lambda))\dot{v}(\lambda)^2\,.
\end{equation}
To avoid technicalities, let us assume that our curve $\gamma$ is either timelike or 
spacelike, with no endpoints. The curve can then be parametrized by proper time 
(resp. proper distance) $\lambda$ ranging from $-\infty$ to $+\infty$, and we assume this 
has been done. We assume furthermore that both $\dot u(\lambda)$ and $\dot v(\lambda)$ 
are bounded away from zero on the parameter range of the curve 
(i.e., greater or equal to some fixed $\eps > 0$), meaning 
that the curve does not become null asymptotically. We also restrict consideration to 
curves that do not ``wiggle'' too rapidly by assuming moreover that all derivatives 
of $\dot u(\lambda)$ and $\dot v(\lambda)$ vanish faster than polynomially.  
Our assumptions imply that the functions $u(\lambda)$ and $v(\lambda)$ can therefore be 
inverted with smooth inverses $\lambda(u)$ resp. $\lambda(v)$, the derivatives of which 
are Schwartz functions.

Let $G$ be a smooth, non-negative Schwartz function. Our assumptions 
then ensure that the smearing functions $G_R(u) = G(\lambda(u))$ and $G_L(v) = G(\lambda(v))$
and consequently $G_R(u) |d\lambda(u)/du|^{-1}$ and $G_L(v) |d\lambda(v)/dv|^{-1}$
are in the Schwartz class. Thus, using the simultaneously sharp QEIs 
for both left- and right-moving stress-energy densities, we obtain the worldline QEI
\begin{multline}
\inf_{\psi\in\DD} \int \langle\rho_\gamma(\lambda)\rangle_\psi
G(\lambda)\,d\lambda \\
= -\frac{c_R}{12\pi} \int\left( \frac{d}{du} 
\sqrt{\frac{G_R(u)}{|d\lambda(u)/du|}}\right)^2 \, du
-\frac{c_L}{12\pi} \int\left( \frac{d}{dv} 
\sqrt{\frac{G_L(v)}{|d\lambda(v)/dv|}}\right)^2 \, dv, 
\label{eq:worldline}
\end{multline}
where the integrands on the right side are set to zero for points such 
that $G_L$ resp. $G_R$ vanish. This bound can be generalized to smooth 
parametrized curves $\gamma^\mu$ satisfying less stringent conditions, but 
we will not go into this here. We only remark that we may also obtain a bound 
for the affinely parametrized left-moving null ray
$u = \lambda, v = {\rm const.}$ for any non-negative $G(\lambda)$ in the Schwartz class. 
In that case, $\rho_\gamma = T_R$ and the worldline bound is given by the QEI bound
for the right-moving stress tensor (with $G_R = G$) given in our theorem.
A similar statement holds of course also for the right moving light 
ray. In general, therefore, averages of the null-contracted stress-energy density
$T_{\mu\nu}k^\mu k^\nu$ are bounded below along an affinely parametrised null
line with tangent $k^\mu$. As noted in~\cite{Flanagan02}, no
other component of the stress tensor can be bounded below along such a
curve because all other components involve $T_R$ or $T_L$ evaluated at a
single point and therefore not averaged.

For the case of a static worldline parametrized by proper time,
$\gamma^0 = x^0, \gamma^1 = x^1 = {\rm const.}$, we find
\begin{equation}
\inf_{\psi\in\DD} \int \langle T_{00}(x^0, x^1)\rangle_\psi G(x^0)\,dx^0
= -\frac{c_L+c_R}{12\pi} \int \left( \partial_0 \sqrt{G(x^0)} \right)^2\,dx^0
\label{eq:staticworldline}
\end{equation}
which reduces to Flanagan's bound~\cite{Flan} for the massless scalar field
($c_L=c_R=1$) and Vollick's bound~\cite{Vollick} for the massless (complex)
Dirac field, which also has $c_L=c_R=1$. [The Majorana field
has $c_L=c_R=1/2$ and a correspondingly tighter bound.]

It is worth noting a feature of conformal quantum field theories in two dimensions: 
namely that one can obtain a (nontrivial) worldline quantum energy
inequality even along spacelike or null curves. This can be traced back to the fact that 
one is free to interchange the role of space and time in two-dimensional conformal field theories
(by ``turning Minkowski space on its side'') as far as the stress-tensor is concerned. Neither is possible in any other 
dimension~\cite{FHR, FewsterRoman03} (even for free scalar fields), nor for 
non-conformally invariant field theories in two dimensions. In those cases, we expect however that 
there still hold bounds for {\em spacetime} averages of the stress tensor, 
to which we now turn.

\subsubsection{Worldvolume bounds}

Let $f^{\mu\nu}$ be a smooth tensor field whose components (with respect
to global inertial coordinates) are Schwartz class. 
Then
\begin{equation}
\int T_{\mu\nu}f^{\mu\nu}(x^0,x^1) \,dx^0dx^1 = \int T_R(u) F_R(u)\,du + \int T_L(v) F_L(v)\,dv  
\label{eq:Tfmunu}
\end{equation}
where the {\em null averages} $F_L$ and $F_R$ are given by
\begin{equation}
F_R(u) = \int f^{uu}(u,v)\,dv\,,\qquad F_L(u) = \int f^{vv}(u,v)\,du
\end{equation}
with $f^{uu}$, $f^{vv}$ appropriate components in
$(u,v)$-coordinates, related to the components in $(x^0,x^1)$ coordinates by
\begin{eqnarray}
f^{uu} &=& f^{00} + f^{11} -f^{01} -f^{10} \nonumber\\
f^{vv} &=& f^{00} + f^{11} +f^{01} +f^{10}\,.
\end{eqnarray}
If $f^{\mu\nu}$ has nonnegative null averages\footnote{This follows of course in particular if 
$f^{\mu\nu}$ satisfies the conditions $f^{uu}, f^{vv} \ge 0$ pointwise.}, 
then we have the worldvolume QEI
\begin{multline}
\inf_{\psi\in\DD}\int \langle T_{\mu\nu}f^{\mu\nu}(x^0, x^1)\rangle_\psi \,dx^0dx^1 \\
= -\frac{c_L}{12\pi} \int \left(\frac{d}{dv} \sqrt{F_L(v)}\right)^2 \, dv 
-\frac{c_R}{12\pi} \int \left(\frac{d}{du} \sqrt{F_R(u)}\right)^2 \,
du\,,
\label{eq:wvolQEI}
\end{multline}
where the integrands on the right side are as usual defined to be zero for 
points $u$ (resp., $v$) where $F_L(u)$ (resp., $F_R(v)$) vanishes. In
particular, if $s^\mu$ and $t^\nu$ are Schwartz-class timelike vector
fields, $f^{\mu\nu}=s^\mu t^\nu$ obeys the above condition and so we
obtain a quantum dominated energy inequality (QDEI). 

\subsubsection{Moving mirrors and boundary CFT}

As a variation on the foregoing results, let us consider a moving mirror
model, with central charge $c$, living in the portion $v>p(u)$ of Minkowski space, where $u\mapsto
p(u)$ lifts to some $\rho\in\tDiffS$. As described in
Sec.~\ref{sect:axioms}, the left- and right-moving components of the
stress-energy density are given in terms of a single field $T$ by the relations $T_L(v)=T(v)$,
$T_R(u)=U(\rho)T(u)U(\rho)^{-1}$. If $f^{\mu\nu}$ is a smooth tensor
field compactly supported in $v>p(u)$, then Eq.~\eqref{eq:Tfmunu} and the
transformation law~\eqref{eq:LMv} entail
\begin{equation}
\int T_{\mu\nu}f^{\mu\nu}(x^0, x^1) \,dx^0dx^1 
=\int T(v)G(v)\,dv -\frac{c}{24\pi}\int \{p,u\}F_R(u)\,du\,,
\end{equation}
where 
\begin{equation}
G(v) = F_L(v)+p'(p^{-1}(v))F_R(p^{-1}(v))\,,
\end{equation}
and an obvious change of variables has also been employed. Thus we have
the modified worldvolume QEI
\begin{multline}
\inf_{\psi\in\DD}\int \langle T_{\mu\nu}f^{\mu\nu}(x^0, x^1)\rangle_\psi \,dx^0dx^1 \\
= -\frac{c}{12\pi} \int \left(\frac{d}{dv} \sqrt{G(v)}\right)^2 \, dv 
-\frac{c}{24\pi}\int \{p,u\}F_R(u)\,du\,,
\label{eq:mirrorQEI}
\end{multline}
in which the last term relates to the stress-energy density created by
the motion of the mirror.

If the support of $f^{\mu\nu}$ is such that the supports of $F_L$ and
$F_R\circ p^{-1}$ (i.e., the two `null projections' of
$f^{\mu\nu}$ onto the mirror trajectory) are disjoint, the first term in
the above bound splits into terms involving $F_L$ and $F_R$ separately.
The term in $F_R$ may be recombined with the final term in
Eq.~\eqref{eq:mirrorQEI}, leading to the same overall result as
in Eq.~\eqref{eq:wvolQEI}. This is to be expected on grounds of
locality, as measurements in (a diamond neighbourhood of) the support of
$f^{\mu\nu}$ should be unaware of the presence of the boundary.
(See also~\cite{RehrenLongo} for a detailed discussion of boundary CFT
in which these ideas also appear.)

\subsubsection{Unweighted averages}

Finally, we discuss unweighted averages of the stress-energy tensor
along portions of a worldline $\gamma$. First, let us note that, if $\gamma$ is
an infinite straight line (with $\dot{u}$ and
$\dot{v}$ constant) then
\begin{equation}
 \int \langle\rho_\gamma(\lambda)\rangle_\psi \,d\lambda \ge 0
\end{equation}
for all $\psi\in\DD$, because the left-hand side is simply a weighted
sum of $P_L$ and $P_R$ with positive coefficients. Accordingly, 
conformal field theories in Minkowski space obey the averaged
weak energy condition, and the averaged null energy condition. 

However, unweighted averaging along a bounded, or even semi-infinite, portion of
such a worldline leads to very different results. For simplicity, we
consider a theory with only one independent component of stress-energy,
and averaging over $(-\infty,0)$, but it is easy to extend these arguments. We begin by
constructing a particular family of states as follows. Let $f\in\CoinX{(-1,1)}$ obey
$f(v)\ge -1$, $\int f(v)\,dv=0$, and suppose $f$ is not identically zero
on $(-1,0)$. Then the map $v\mapsto V(v)$ defined by 
\begin{equation}
V(v) =v+ \int_{-1}^v f(v')\,dv'
\end{equation}
is a diffeomorphism of the line which lifts to some
element $\rho\in\tDiffS$ (as it agrees with the identity outside a
compact interval). If $f$ obeys, additionally, 
\begin{equation}
-1\le \frac{d^2}{dv^2} \frac{1}{\sqrt{1+f(v)}} \le 0
\end{equation}
for $v\in(-1,0)$, then $\{V,v\}\ge 0$ on this interval, and no conflict
need arise with our previous assumptions because the left-hand
inequality ensures that $\int_{-1}^0 f(v)dv<1$. Moreover $\{V,v\}$ must be
strictly positive on some open subset of $(-1,0)$, since $f$ is not
identically zero there. Owing to the identity
\begin{equation}
\int \frac{\{V,v\}}{\sqrt{V'(v)}} \,dv = -2\int
\frac{d^2}{dv^2}\frac{1}{\sqrt{V'(v)}}\,dv =0\,,
\end{equation}
it follows that $\{V,v\}$ is strictly negative on some open subset of $(0,1)$
(note that $\{V,v\}$ is supported in $(-1,1)$). 

With the above assumptions in force, we may use the resulting
diffeomorphism to create a vector state $\psi=U(\rho)^{-1}\Omega$ by acting on the vacuum.
The corresponding energy density, 
\begin{equation}
\langle T(v)\rangle_\psi = -\frac{c}{24\pi}\{V,v\}\,,
\end{equation}
is smooth and compactly supported in
$(-1,1)$, nonpositive for $v\le 0$, and strictly negative (resp.,
positive) on some open subset of $(-1,0)$ (resp., $(0,1)$). In
particular, 
\begin{equation}
\int_{-\infty}^0 \langle T(v)\rangle_\psi = -\frac{c}{24\pi}\int_{-\infty}^0
\{V,v\}\,dv <0\,.
\end{equation}
We now consider the family of states obtained by scaling $\psi$, namely
$\psi_\lambda=U(D_\lambda)^{-1}\psi$, for which
\begin{equation}
\int_{-\infty}^0 \langle T(v)\rangle_{\psi_\lambda}\,dv
= -\frac{c\lambda}{24\pi}\int_{-\infty}^0
\{V,v\}\,dv \to-\infty
\end{equation}
as $\lambda\to\infty$. The reason for this is that
the negative energy density becomes more and more sharply
peaked near zero under the dilations, with magnitude growing like
$\lambda^2$ and support shrinking as $\lambda^{-1}$. Thus we have
shown explicitly that sharp averages
of the stress-energy density are not subject to QEI restrictions. A
related result holds for general quantum fields with mass-gap in two
dimensions, as shown by Verch (Prop.~3.1 of~\cite{VerchANEC}).

However, there is no contradiction between this observation and the QEIs proved
above. An average taken against a weight function
$G\in\CoinX{-\infty,0}$ in the vector states $\psi_\lambda$ would in fact tend
to zero as $\lambda\to\infty$ because the negative peak eventually
leaves the support of $G$. If one used a weight function which did not
vanish at the origin, its support would spill over into the right-hand half
line and sense the energy density there. However, the family of states $\psi_\lambda$ also has an
increasingly sharply peaked {\em positive} energy density within the
interval $(0,\lambda^{-1})$, which must at least compensate for the
negative contribution (because $\int \langle T(v)\rangle_{\psi_\lambda}\,dv$
is nonnegative). It is the competition between these two differently weighted contributions 
which permits the QEI to hold. 

To emphasise the point, let us consider averages over half the light-ray, but with a
smoothed-off end. Let $G$ be a nonnegative, smooth and compactly
supported function, which equals unity in a neighbourhood of the origin.
Define a sequence of smooth functions $G_n$ by 
\begin{equation}
G_n(v) = \vartheta(-v)G(v/n)+\vartheta(v)G(v)\,,
\end{equation}
where $\vartheta$ is the Heaviside function (and we take
$\vartheta(0)=1/2$). As $n\to\infty$, these functions approach 
$H(v)=\vartheta(-v)+\vartheta(v)G(v)$. Now for any non-zero $\psi\in\DD$ we have 
\begin{align}
\int \langle T(v)\rangle_\psi G_n(v)\,dv &\ge -\frac{c}{12\pi}\int
\left(\frac{d}{dv}\sqrt{G_n(v)}\right)^2\,dv \\
&=-\frac{c}{12\pi}\int \left[\frac{\vartheta(-v)}{n}+\vartheta(v)\right]
\left(\frac{d}{dv}\sqrt{G(v)}\right)^2\,dv 
\end{align}
for each $n$. Taking $n\to\infty$ and using the fact that $\langle
T(v)\rangle_\psi$ decays as $O(v^{-4})$ (by the remark following
Eq.~\eqref{eq:PasintT}) we obtain
\begin{equation}
\int \langle T(v)\rangle_\psi H(v)\,dv \ge -\frac{c}{12\pi}\int_0^\infty
\left(\frac{d}{dv}\sqrt{G(v)}\right)^2\,dv 
\end{equation}
for arbitrary $\psi\in\DD$. As expected, the bound
depends only on the way the averaging is rounded-off.

\section{Highest-weight Virasoro representations}\label{sect:virasoro}

In this section, we describe how CFT models satisfying our axioms may be constructed
by taking direct sums of unitary, highest-weight representations of the
Virasoro algebra. In particular, this demonstrates that our QEI applies
to so-called minimal models and to rational conformal field theories. As
part of our discussion we will need to consider the unitary multiplier
representations of $\tDiffS$ carried by any
such Virasoro representation; in particular, we need to show that the
representation can be normalised so that the multiplier is of the Bott
form assumed in Axiom A.1. We have not found this elsewhere in the literature.

\subsection{Highest-weight representations of the Virasoro algebra}

We recall that the Virasoro algebra is generated by elements
$L_n$ ($n\in\ZZ$) and a central element $\kappa$, obeying the relations
\begin{equation}
[L_m, L_n] = (m-n)L_{m+n} +\frac{1}{12}m(m^2-1)\delta_{m+n,0}\kappa\qquad(m,n\in\ZZ)
\label{eq:Vir}
\end{equation}
and $[\kappa,L_m]=0$ for all $m\in\ZZ$. A unitary highest-weight representation
amounts to the specification of a pair $(c,h)$ of real constants, a Hilbert space
$\HH_{(c,h)}$, a dense domain
$\DD_0\subset\HH_{(c,h)}$, a vector $\ket{h}\in\DD_0$, and
operators $L_n$ ($n\in\ZZ$) defined on $\DD_0$ such that
\begin{enumerate}
\item $L_0\ket{h}=h\ket{h}$ and $L_n\ket{h}=0$ for $n>0$. 
\item $\DD_0$ coincides with the set of vectors obtained from $\ket{h}$ by
acting with polynomials in the $L_n$ with $n<0$ (including the trivial
polynomial $\II$).
\item $L_n^*=L_{-n}$ on $\DD_0$ and Eq.~\eqref{eq:Vir} holds as an identity on
$\DD_0$ with $\kappa=c\II$. 
\end{enumerate}
Such representations are irreducible; moreover, the `highest weight'
$(c,h)$ is restricted to particular values first classified
in~\cite{FQS84,GKO}. (See, e.g., Theorems 6.17(3) and 6.13 in~\cite{Schott}.)
However, we will not need the precise details of this
classification beyond the fact that both $c$ and $h$ are nonnegative,
which follows immediately from the observation that
$0\le \|L_{-n}\ket{h}\|^2 = 2nh+n(n^2-1)c/12$ for all $n\ge 1$ as a
consequence of Eq.~\eqref{eq:Vir}. 

In the course of our analysis, we will need more detailed information on
the domain of definition of the $L_n$ and various other operators. 
Our first observation is that, by virtue of the Virasoro relations, 
$\DD_0$ contains an orthonormal basis of $L_0$-eigenvectors. Indeed, this
follows by the Gram--Schmidt process applied to vectors of the form
$L_{-n_1}L_{-n_2}\cdots L_{-n_k}\ket{h}$ (for $n_1,\ldots,n_k>0$), which are $L_0$-eigenvectors
with eigenvalue $h+n_1+n_2+\cdots + n_k$. Thus $L_0$ is essentially self-adjoint on
$\DD_0$ and we will use $L_0$ from now on to
denote the unique self-adjoint extension of this operator, writing
$D(L_0)$ for its domain. The above remarks also show that $L_0$ is a
positive operator, with spectrum contained in $h+\NN_0$ and
finite-dimensional eigenspaces. Secondly,
estimates obtained by Goodman and Wallach~\cite{GW}\footnote{See~\cite{BSM90}
for related bounds.} entail that
\begin{equation}
\|L_n\psi\|\le C(1+|n|)^{3/2}\|L_0\psi\|
\label{eq:GWestimate}
\end{equation}
for all $\psi\in\DD_0$ and $n\in\ZZ$, where the constant $C$ is determined
by the central charge and is independent of both
$\psi$ and $n$. Accordingly, the $L_n$ may be extended uniquely to
$D(L_0)$, and we now use $L_n$ to denote these extensions. The relation
$L_n=L_{-n}^*$ continues to hold, and the Virasoro relations hold as
identities on $D(L_0^2)$. A further consequence is that the formula
\begin{equation}
\Theta(z) = -\frac{1}{2\pi} \sum_{n\in\ZZ} z^{-n-2}L_n \,,
\label{eq:stressdef}
\end{equation}
defines $\Theta(\cdot)\psi$ as a vector-valued distribution on $C^\infty(S^1)$ 
for each $\psi\in D(L_0)$.  Furthermore, 
\begin{equation}
\Theta(f)^* = \Theta(\Gamma f) 
\end{equation}
on $D(L_0)$ for $f\in C^\infty(S^1)$. In particular, if
$\Gamma f=f$ (i.e., $f\in C^\infty_\Gamma(S^1)$) then $\Theta(f)$ is symmetric on $D(L_0)$ and an
application of Nelson's commutator theorem (Theorem X.37
in~\cite{RSii}\footnote{In the notation of~\cite{RSii}, set
$A=\Theta(f)$, $N=L_0+\II$ and $D=\DD_0$, for example.}) shows that
$\Theta(f)$ is essentially self-adjoint on any core of $L_0$. Henceforth
we will use $\Theta(f)$ to denote the unique self-adjoint extension.
It is easy to verify that the $\Theta(f)$'s defined in this way obey the
commutation relations Eq.~\eqref{eq:TVirasoro} on $D(L_0^2)$. 

Finally, let us define the space $\HH^\infty$ to be the intersection
$\HH^\infty = \bigcap_{n\in\NN_0} D(L_0^n)$, equipped with the Fr\'echet
topology induced by the seminorms $\psi\mapsto\|L_0^n \psi\|$ ($n\in\NN_0$).
As $\DD_0\subset D(L_0^n)$ for each $n$, it follows that $\HH^\infty$ is
dense in $\HH$ and is a core for $L_0$.

\subsection{Integration to a unitary representation of $\protect\tDiffS$}

We now need to demonstrate that $\Theta$ generates a unitary multiplier
representation of $G=\tDiffS$ as in Axiom A.1 and
Eq.~\eqref{eq:Tfdef}. The relevant results are all present in the
literature, but do not appear to have been assembled in this form before. Explicit
control of the multiplier is necessary when we come to assemble Virasoro
representations to form more general CFT models below: the direct sum of
two projective representations is not generally a projective representation!

Let $\cU_{(c,h)}$ be the group of unitary operators on $\HH_{(c,h)}$ and
let $\PU_{(c,h)}$ be the projective unitary group (i.e., unitaries
modulo phases) $\PU_{(c,h)}=\cU_{(c,h)}/\TT$. In the following we distinguish unitary
multiplier representations (which take values in $\cU_{(c,h)}$) from
projective unitary representations (which take values in $\PU_{(c,h)}$). 
As shown by Goodman and Wallach~\cite{GW}\footnote{In fact~\cite{GW}
addresses $\Diff_+(S^1)$ rather than its universal cover.} and Toledano Laredo~\cite{TL},
$\HH_{(c,h)}$ carries a projective unitary representation $\UU$ of $G$, so the
remaining problem is to assign phases in such a way that Axiom A.1 and
Eqs.~\eqref{eq:multiplier} and~\eqref{eq:Tfdef} are satisfied.

It is helpful (and standard) to rephrase this problem in a geometric
fashion. Let $\Ghat$ be the subgroup of $G\times\cU_{(c,h)}$ defined by 
\begin{equation}
\Ghat = \{(g,V)\in G\times\cU_{(c,h)}: \UU(g) = p(V)\}
\end{equation}
where $p:\cU_{(c,h)}\to \PU_{(c,h)}$ is the quotient map. As shown in 
Proposition~5.3.1 of~\cite{TL} $\Ghat$ is a central extension of $G$ by $\TT$
which may be given the structure of a Lie group. In particular, it is a
smooth principal $\TT$-bundle over $G$ (with projection $\pi(g,V)=g$):
the problem of assigning local (respectively, global) phases to $\UU$ is
then equivalent to selecting a local (resp., global) section of $\Ghat$.

The local problem was addressed by Toledano Laredo in the course of
proving the result just mentioned. He showed 
that phases can be assigned to $\UU$ in a neighbourhood $N$ of $\id$ to
provide a local unitary multiplier representation $U_{\rm loc}$ of $G$ 
so that (i) the map $(g,\psi)\mapsto U_{\rm loc}(g)\psi$ is smooth from
$N\times \HH^{\infty}$ to $\HH^{\infty}$ and (ii) for each $f\in
C^\infty_\Gamma(S^1)$ and each $\psi\in\HH^{\infty}$,
\begin{equation}
\left.\frac{d}{ds}U_{\rm loc}(e_{f}(s))\psi\right|_{s=0} =i
\Theta(f)\psi
\label{eq:TLsinfini}
\end{equation}
where $s\mapsto e_{f}(s)$ is a smooth curve in $G$ with $e_{f}(0)=\id$ and
$\dot{e}_{f}(0)=\vf$, the corresponding vector field to $f$. 
[These curves, and $U_{\rm loc}$, are determined by a choice of
coordinates near $\id$.] By (i) we may replace $e_{f}(s)$ by $\exp s \vf$ in
Eq.~\eqref{eq:TLsinfini}, so $U_{\rm loc}$ obeys Eq.~\eqref{eq:Tfdef}
and provides a local solution to our problem. A further consequence of
(i) is that $U_{\rm loc}$ is strongly continuous on $\HH$, because $\HH^\infty$
is dense in $\HH$ and the $U_{\rm loc}(g)$ have unit operator norms. 
Toledano Laredo also uses
$U_{\rm loc}$ to show that the Lie algebra cocycle of $\Ghat$ is
cohomologous to $c\omega$, where $\omega$ is the Virasoro cocycle of Eq.~\eqref{eq:Vircocycle}.

The global assignment of phases is achieved by the following result. 
\begin{Prop} 
There is a global smooth section $g\mapsto (g,U_{(c,h)}(g))$ of
$\Ghat$ such that $g\mapsto U_{(c,h)}(g)$ is a strongly continuous unitary multiplier representation
of $G$ leaving $\HH^\infty$ invariant and
obeying
\begin{equation}
U_{(c,h)}(g)U_{(c,h)}(g') = e^{ic\widetilde{B}(g,g')}U_{(c,h)}(gg')\qquad (g,g'\in G)\,.
\end{equation}
Moreover, if $\vf\in\Vect_\RR(S^1)$ is the vector field corresponding to
$f\in C^\infty_\Gamma(S^1)$ then $D(\Theta(f))$ consists precisely of
those $\psi\in\HH$ for which $s\mapsto U_{(c,h)}(\exp s\vf)\psi$ is
differentiable, and we have
\begin{equation}
\left.\frac{d}{ds}U_{(c,h)}(\exp s\vf)\psi\right|_{s=0} =i \Theta(f)\psi
\label{eq:infini}
\end{equation}
for such $\psi$.
\end{Prop}
{\noindent\em Remark:} As discussed in Sect.~\ref{sect:LieGroupStructure}, $G$ is diffeomorphic to a convex subset
of the Fr\'echet space $C^\infty_{2\pi}(\RR;\RR)$. Accordingly, Poincar\'e's lemma (see e.g., Lemma 3.3
in~\cite{Neeb}) entails that $G$ has trivial cohomology groups
$H^k(G;\RR)$. In consequence, $H^2(G;\ZZ)$, which classifies the smooth
principal $\TT$-bundles over $G$ (see Sect.~4.5 in~\cite{PS86}) is also
trivial, so $\widehat{G}$ is isomorphic to $G\times\TT$ as a smooth
manifold and therefore admits smooth global sections.\\
{\noindent\em Proof:} By Proposition~4.2 in~\cite{Neeb} $\Ghat$ may be
described in terms of a group 2-cocycle mapping $G\times G$ to $\TT$
which is smooth near $(\id,\id)$. Because $G$ is simply connected, the
equivalence class of group cocycles describing $\Ghat$ is fixed by the infinitesimal
class of $c\omega$ (see, e.g., the long exact sequence of Theorem 7.12
in~\cite{Neeb}) and therefore includes the Bott cocycle 
$\Omega_c(g,g')=e^{ic\widetilde{B}(g,g')}$ for central
charge $c$. Let $g\mapsto (g,V(g))$ be any smooth global section of
$\Ghat$ and define the corresponding (everywhere smooth) cocycle $m:G\times G\to\TT$ by
$V(g)V(g')=m(g,g')V(gg')$. Since $m$ and $\Omega_c$ are cohomologous there
exists $\mu:G\to \TT$, smooth near $\id$, such that
\begin{equation}
m(g,g') = \Omega_c(g,g') \frac{\mu(gg')}{\mu(g)\mu(g')}\,.
\end{equation}
As both $m$ and $\Omega_c$ are smooth it follows that $\mu$ is everywhere
smooth; the required global section is given by $U_{(c,h)}(g)=\mu(g)V(g)$. 

Near the identity, we must have $U_{(c,h)}(g)=e^{i\nu(g)}U_{\rm loc}(g)$
for some smooth $\nu:N\to\RR$. It follows that $U_{(c,h)}$ is strongly continuous on
$\HH$ and has well-defined generators $\Xi(f)$ given on $\HH^\infty$ by 
\begin{equation}
i\Xi(f)\psi = \left.\frac{d}{ds}U_{(c,h)}(\exp s\vf)\psi\right|_{s=0}\,,
\end{equation}
and obeying $\Xi(f) = \Theta(f) + \alpha(f)\II$ (on $\HH^\infty$) where
$\alpha(f)=\nu'_{\id}(\vf)$ is continuous and linear in $f\in C^\infty_\Gamma(S^1)$ because $\nu$ is smooth.
By Prop.~\ref{prop:transform}, applied to $U_{(c,h)}$ and $\HH^\infty$, 
the generators $\Xi$ obey the same algebraic relations on $\HH^\infty$
as the $\Theta$'s on $\HH^\infty$. In particular they obey 
Eq.~\eqref{eq:TVirasoro}, from which it follows that $\alpha(fg'-f'g)=0$ for all
$f,g\in C^\infty_\Gamma(S^1)$. 
It is now straightforward to show that $\alpha$ vanishes
on a basis for $C^\infty_\Gamma(S^1)$ and hence identically.
Accordingly, Eq.~\eqref{eq:infini} holds for $\psi\in\HH^\infty$ and, in
particular, on $\DD_0$. Now, the argument of footnote~\ref{fn:stone}
above guarantees that the left-hand side of Eq.~\eqref{eq:infini}
defines a self-adjoint operator whose domain consists precisely of those
$\psi$ for which the derivative exists. As this operator agrees with
$\Theta(f)$ on a core, it must in fact be $\Theta(f)$.  $\square$

We have thus established that the stress-energy density in a unitary
highest-weight Virasoro representation is the
infinitesimal generator of a unitary multiplier representation of
$\tDiffS$ with the Bott cocycle. Thus $\HH_{(c,h)}$ and $U_{(c,h)}$ satisfy
axiom A.1 of Sect.~\ref{sect:axioms}. Moreover the algebraic relations
Eqs.~\eqref{eq:LM_smeared} and~\eqref{eq:TVirasoro} hold when applied to vectors in $\HH^\infty$.
Let us observe that it is {\em not} the case that
\begin{equation}
U_{(c,h)}(\exp s\vf) = e^{is\Theta(f)} \qquad \hbox{(FALSE)}
\label{eq:false}
\end{equation}
for all $s\in\RR$ and $f\in C^\infty_\Gamma(S^1)$ because the Bott cocycle
does not vanish along all one-parameter subgroups (although it is of
course a coboundary). In passing we mention
that Goodman and Wallach~\cite{GW} appear to claim that their unitary multiplier
representation of $\Diff_+(S^1)$ can be normalised in such a way that
Eq.~\eqref{eq:false} holds. However, this cannot be true, as it is not possible
for exponentiations of ${\mathfrak sl}(2,\RR)$ representations with
noninteger highest weight. 

Turning to axiom A.2, we note that representations with $h\not=0$
do not contain a vacuum vector invariant under $U_{(c,h)}|_{\widetilde{\hbox{\scriptsize
M\"ob}}}$, because this representation of $\widetilde{\Mob}$ is
generated by $L_0$ and linear combinations of $L_{\pm 1}$, and we know that $\spec(L_0)\subset
h+\NN_0$. If $h=0$ the highest-weight vector $\ket{0}$ is indeed the
unique invariant vector, as required by axiom A.2.\footnote{For arbitrary highest-weight $h$, the
highest-weight vector $\ket{h}$ obeys
$L_0\ket{h}=h\ket{h}$, $L_1\ket{h}=0$, $\|L_{-1}\ket{h}\|^2=2h$. The
assertion follows on taking $h=0$.} We will return to this when
constructing more general CFT models.  

Continuing with general highest-weight Virasoro representations, 
Axiom A.3 clearly holds, because the generator of rotations $H=L_0$ is positive.
To check the remaining axioms, we construct a new $U_{(c,h)}$-invariant domain,
\begin{equation}
\DD_{(c,h)}={\rm span}\,\bigcup_{g\in G} U_{(c,h)}(g)\DD_0\,,
\label{eq:Ddef}
\end{equation}
(i.e., finite linear combinations of vectors of form $U_{(c,h)}(g)\psi$
for $g\in G$, $\psi\in\DD_0$). It is clear that $\DD_{(c,h)}$ lies within
$\HH^\infty$, as $\DD_0\subset\HH^\infty$ and $\HH^\infty$ is
$U$-invariant. Thus $\DD_{(c,h)}\subset D(L_0)\subset D(\Theta(f))$ for each $f\in
C^\infty(S^1)$, verifying axiom B.1 (apart from the statement concerning
the vacuum, which holds if and only if $h=0$ for the vector $\ket{0}$).
Moreover, the comment after Eq.~\eqref{eq:stressdef} shows that $\Theta(\cdot)\psi$
is a vector valued distribution on $C^\infty(S^1)$ for each $\psi\in\DD_{(c,h)}$.
Accordingly, $\HH_{(c,h)}$, $U_{(c,h)}$ and $\DD_{(c,h)}$ satisfy axiom B.2. 

We also wish to see that expectation values of $\langle
\Theta(z)\rangle_\psi$ for $\psi\in\DD_{(c,h)}$ are smooth. This can be
verified directly for $\psi\in\DD_0$, in which case the expectation
values are polynomial in $z$ and $z^{-1}$; the extension to $\DD_{(c,h)}$ then
follows from the transformation law Eq.~\eqref{eq:LM_smeared} (which
holds on $\HH^\infty$ and hence on $\DD_{(c,h)}$). Thus axiom B.3 holds. 

To summarise: we have established that $\HH_{(c,h)}$, $\DD_{(c,h)}$ and
$U_{(c,h)}$ obey all the axioms for a CFT on $S^1$ except those relating
to the vacuum state; all the axioms are obeyed if $h=0$. 

\subsection{CFT models obeying the axioms}\label{sect:models}

It is now easy to construct a large class of theories obeying
our axioms, simply by taking direct sums of Virasoro representations. 
Starting with CFTs with a single component of stress-energy, we may take, for example,
\begin{equation}
\HH=\bigoplus_{k=0}^K \HH_{(c,h_k)}\,,\qquad U=\bigoplus_{k=0}^K U_{(c,h_k)}\,,
\end{equation}
where $0\le  K\le \infty$ and $0=h_0<h_1\le h_2\le h_3\cdots$ with each
$(c,h_k)$ an allowed highest weight for a unitary representation of the
Virasoro algebra. Here we
take $\DD$ to be the space of vectors in $\HH$ with only finitely many
nonzero components, each belonging to the appropriate $\DD_{(c,h_k)}$,
and set $\Omega = (\ket{0},0,0,\ldots)$. Since we argued in the last 
subsection that the multipliers in each
$U_{(c,h_k)}$ are all equal, their direct sum is also a unitary multiplier
representation with the same multiplier. In addition, by insisting on a
unique summand with $h=0$, we have guaranteed the existence of a unique
vacuum vector. 

In a similar fashion, CFTs with two independent components of stress-energy
may be constructed as direct sums of tensor products of the form
\begin{equation}
\HH = \bigoplus_{k=0}^K \HH_{(c_L,h_{L,k})}\otimes \HH_{(c_R,h_{R,k})}\,,
\end{equation}
in which $0\le K\le\infty$ as before, and we require that
$(h_{L,k},h_{R,k})=(0,0)$ if and only if $k=0$. The vacuum is
$\Omega=(\ket{0}\otimes\ket{0},0,0,\ldots)$ (and is again unique) and the space $\DD$ is
constructed as before. Thus our axioms embrace, and are more general than, minimal
models (for which $K$ is finite) and rational conformal
field theories (for which $K$ may be infinite but the theory is
minimal for an extended algebra, e.g., minimal superconformal
models~\cite{FQS84} or WZW models~\cite{Witten84}).

\medskip

{\noindent\em Acknowledgments:} 
The work of CJF was assisted by 
EPSRC Grant GR/R25019/01 to the University of York. SH was supported by 
NSF Grant PH00-90138 to the University of Chicago, by NSF Grant
PHY0354978 and by funds from the University of California. Part of this research 
was carried out during the 2002~program on Quantum Field Theory in Curved
Spacetime at the Erwin Schr\"odinger Institute, Vienna, and we wish to thank the 
Institute for its hospitality. We have greatly benefited from conversations
with participants of the program, in particular \'E.\'E.~Flanagan,
K.~Fredenhagen and K.-H.~Rehren. CJF would also like to thank
G.W.~Delius and I. McIntosh for 
many illuminating discussions on conformal field theory and infinite-dimensional Lie groups. 

\appendix
\section{Square roots of Schwartz class functions}

In the body of the paper, we use various properties of square roots
of functions in the Schwartz class. The following results are quite
probably known, but are included for completeness.  
Related results, also based on the use of Taylor's theorem, may be found
in e.g. Lemma~1 of~\cite{Glaeser} and p.~86 of~\cite{DimSjo}.

\begin{Lem} \label{Lem:poslem}
Let $G\in\Sch(\RR)$ be nonnegative. Then there exists $M>0$
such that
\begin{equation}
G'(v)^2 \le \frac{4MG(v)}{1+v^2}
\label{eq:sqclaim}
\end{equation}
for all $v$. In particular, $|d/dv\, \sqrt{G(v)}|^2\le M/(1+v^2)$ where
$G(v)\not=0$. 
\end{Lem}
{\noindent\em Proof:} Noting that the result holds trivially if $G\equiv
0$, we now restrict to nontrivial $G$. For $k=0,2$, let $M_k=\sup_{v\in\RR}
|(1+|v|^k)G''(v)|$, observing that each $M_k>0$. If $\epsilon>0$, Taylor's theorem entails that
\begin{equation}
0\le G(v-\epsilon G'(v)) = G(v) - \epsilon G'(v)^2 +
\frac{1}{2}\epsilon^2 G'(v)^2 G''(\eta)
\label{eq:Taylor}
\end{equation}
for some $\eta$ lying between $v$ and $v-\epsilon G'(v)$. We apply this
in two ways. First, for any $v$, we use $G''(\eta)<M_0$ and put
$\epsilon=M_0^{-1}$ to find
\begin{equation}
0 \le G(v) - \epsilon G'(v)^2 +
\frac{1}{2}\epsilon^2 G'(v)^2 M_0 = G(v) - \frac{G'(v)^2}{2M_0}
\end{equation}
so $G'(v)^2\le 2M_0 G(v)$ for all $v$. Second, we observe that
\begin{equation}
2\left|\frac{G'(v)}{v}\right| \le \frac{M_2}{1+(v/2)^2}
\end{equation}
holds for all sufficiently large $|v|$, so setting 
$\epsilon=(1+(v/2)^2)/M_2$ the $\eta$ in Eq.~\eqref{eq:Taylor} obeys
$|\eta|\ge |v|/2$ and we find
\begin{equation}
0 \le G(v) - \epsilon G'(v)^2 +
\frac{\epsilon^2 G'(v)^2 M_2}{2(1+(v/2)^2)} 
= G(v) -\frac{1+(v/2)^2}{2M_2}G'(v)^2
\end{equation}
for all $|v|$ greater than some $v_0>0$. Thus Eq.~\eqref{eq:sqclaim} holds
with $M=\max\{\frac{1}{2}M_0(1+v_0^2),4M_2\}$. $\square$

\begin{Cor} \label{Cor:W1}
Given $0\le G\in\Sch(\RR)$ define
\begin{equation}
\varphi(v) = \left\{\begin{array}{cl} G'(v)/(2\sqrt{G(v)}) &
G(v)\not=0\\ 0 & G(v)=0\,.\end{array}\right.
\end{equation}
Then $\varphi\in L^2(\RR)$ and $\varphi=d/dv \, \sqrt{G}$, where $d/dv$ denotes
the derivative in the sense of distributions. Thus $\sqrt{G}$ belongs to
the Sobolev space $W^1(\RR)$. Furthermore, 
\begin{equation}
\int \varphi(v)^2 \, dv = \lim_{\epsilon\to 0^+} \int \frac{G'(v)^2}{4(G(v)+\epsilon)}\,dv\,.
\end{equation}
\end{Cor}
{\noindent\em Proof:} For $\epsilon>0$ define $G_\epsilon(v) = 
(\sqrt{G(v)+\epsilon}-\sqrt{\epsilon})^2$. Then $\sqrt{G_\epsilon}\to \sqrt{G}$
in $L^2(\RR)$ as $\epsilon\to 0^+$. Moreover
\begin{equation}
\left|\frac{d}{dv}\sqrt{G_\epsilon(v)} - \varphi(v) \right|
= \left|\frac{G'(v)}{2\sqrt{G(v)+\epsilon}}-\varphi(v)\right|
\le 2\left(\frac{M}{1+v^2}\right)^{1/2}\,,
\end{equation}
where $M$ is the constant furnished by Lemma~\ref{Lem:poslem}. (In the
case $G(v)\not=0$, this follows from the triangle inequality; the case
$G(v)=0$ is trivial as we must also have $G'(v)=0$ by Eq.~\eqref{eq:sqclaim}, so the left-hand
side vanishes.)
Since $d/dv \sqrt{G_\epsilon(v)}\to\varphi(v)$ pointwise as $\epsilon\to 0^+$, we deduce that
the convergence occurs in $L^2(\RR)$ by the dominated convergence theorem. Thus
$\varphi=d/dv \sqrt{G}\in L^2(\RR)$. The expression for $\|\varphi\|^2$ is also
proved by dominated convergence. $\square$


\begin{thebibliography}{ZZ}

\bibitem{Alcubierre} M. Alcubierre, `The warp drive: hyper-fast travel within general relativity',
                 Class. Quantum Grav. {\bf 11} (1994) L73-L77. 

\bibitem{BSM90} D. Buchholz and H. Schulz-Mirbach, 
`Haag duality in conformal quantum field theory', Rev. Math. Phys. {\bf 2} (1990), 105--125.

\bibitem{CarpiWeiner} S. Carpi and M. Weiner, `On the uniqueness of
diffeomorphism symmetry in conformal field theory', {\tt math.OA/0407190}.

\bibitem{DimSjo}  M. Dimassi and J. Sj\"ostrand, 
                  {\em Spectral asymptotics in the semi-classical limit}, LMS Lecture Note Series 268, 
                  (Cambridge University Press, Cambridge, 1999).

\bibitem{EFV} S.P. Eveson, C.J. Fewster and R. Verch, `Quantum Inequalities in Quantum
Mechanics'. Ann. Henri Poincar\'e {\bf 6} (2005), 1--30. 

\bibitem{AGWQI} C.J. Fewster, `A general worldline quantum inequality',
                Class. Quantum Grav. {\bf 17} (2000) 1897--1911.

\bibitem{Lisbon} C.J. Fewster, `Energy Inequalities in Quantum Field Theory',
Expanded version of a contribution to appear in the Proceedings of the XIV International Conference on
Mathematical Physics, Lisbon, 2003. {\tt math-ph/0501073}

\bibitem{FewsterEveson} C.J. Fewster and S.P. Eveson, `Bounds on negative energy densities in flat spacetime',
                Phys. Rev. D58 (1998) 084010.

\bibitem{FewsterMistry} C.J. Fewster and B. Mistry, `Quantum Weak Energy Inequalities
for the Dirac field in Flat Spacetime', Phys. Rev. D {\bf 68} (2003) 105010. 

\bibitem{FewsterPfenning} C.J. Fewster and M.J. Pfenning, 
                          `A Quantum Weak Energy Inequality for spin-one fields in
curved spacetime', J. Math. Phys. {\bf 44} (2003) 4480--4513.

\bibitem{FewsterRoman03} C.J. Fewster and T.A. Roman, `Null energy conditions in quantum field theory',
Phys. Rev. D {\bf 67}, (2003) 044003.

\bibitem{FTi}   C.J. Fewster and E. Teo, `Bounds on negative energy densities in static space-times',
                Phys. Rev. D{\bf 59} (1999) 104016.

\bibitem{FVdirac} C.J. Fewster and R. Verch, `A Quantum Weak Energy Inequality for Dirac fields in curved
spacetime', Commun. Math. Phys. {\bf 225}, (2002) 331--359. 

\bibitem{Flan}   \'E.\'E. Flanagan, `Quantum inequalities in two-dimensional Minkowski spacetime',
                 Phys. Rev. D {\bf 56} (1997) 4922--4926. 

\bibitem{Flanagan02} \'E.\'E. Flanagan, `Quantum inequalities in two dimensional curved spacetimes',
Phys. Rev. D {\bf 66} (2002) 104007.

\bibitem{Ford78} L.H. Ford, `Quantum coherence effects and the second
law of thermodynamics',
                 Proc. R. Soc. Lond. A {\bf 364} (1978) 227--236. 

\bibitem{FHR}     L.H. Ford, A. Helfer, and T.A. Roman, `Spatially averaged quantum
inequalities do not exist in four-dimensional spacetime', 
                  Phys. Rev. D {\bf 66} (2002) 124012.

\bibitem{FRworm} L.H. Ford and T.A. Roman, `Quantum field theory constrains traversable wormhole geometries'
                Phys. Rev. D {\bf 53} (1996) 5496--5507.

\bibitem{FRqis}  L.H. Ford and T.A. Roman, `Restrictions on negative energy density in flat spacetime',
                 Phys. Rev. D {\bf 55} (1997) 2082--2089.
      
\bibitem{FullingDavies} S.A. Fulling and P.C.W. Davies, `Radiation from
a moving mirror in two dimensional space-time: conformal anomaly', Proc.
R. Soc. Lond. A {\bf 348} (1976) 393--414. 

\bibitem{FST} P. Furlan, G.M. Sotkov and I.T. Todorov, `Two-dimensional
conformal field theory',
              Riv. Nuovo Cimento, {\bf 12} (1989) No.~6, 1--202.

\bibitem{FQS84} D. Friedan, Z. Qiu and S. Shenker, `Conformal
invariance, unitarity and critical exponents in two dimensions',
                Phys. Rev. Lett., {\bf 52} (1984) 1575--1578.

\bibitem{Glaeser} G. Glaeser, 
                  `Racine carr\'ee d'une fonction diff\'erentiable',
                  Ann. Inst. Fourier, Grenoble, {\bf 13} (1963) 203--210.

\bibitem{GKO} P. Goddard, A. Kent, and D. Olive,
`Unitary representations of the Virasoro and super-Virasoro algebras',
Commun. Math. Phys., {\bf 103} (1986) 105--119.

\bibitem{GW}      R. Goodman and N.R. Wallach, 
                  `Projective unitary positive-energy representations of $\Diff(S^1)$', 
                  J. Funct. Anal. {\bf 63} (1985), 299--321.


\bibitem{Hamilton} R.S. Hamilton, `The inverse function theorem of Nash and
Moser', Bull. Amer. Math. Soc., {\bf 7} (1982) 65.

\bibitem{Koster_observables} S. K\"oster, `Conformal transformations as observables', 
                             Lett. Math. Phys., {\bf 61} (2002) 187--198.

\bibitem{Koster_absence} S. K\"oster, `Absence of stress energy tensor in
CFT${}_2$ models', {\tt math-ph/0303053}.

\bibitem{KRY}    J. Kupsch, W. R\"uhl and B.C. Yunn, `Conformal invariance of quantum fields in two-dimensional
space-time',
                 Ann. Phys., {\bf 89} (1975) 115--148.

\bibitem{RehrenLongo} R. Longo, K.-H. Rehren, `Local fields in boundary conformal
QFT', Rev. Math. Phys., {\bf 16} (2004) 909--960.

\bibitem{Lu76}   M. L\"uscher, `Operator product expansions on the
vacuum in conformal quantum field theory in two spacetime dimensions'
                 Commun. Math. Phys., {\bf 50} (1976) 23--52.

\bibitem{LusMack} M. L\"{u}scher and G. Mack, `The energy momentum
tensor of a critical quantum field theory in 1+1 dimensions' unpublished
manuscript, 1976.

\bibitem{Mack} G. Mack, `Introduction to conformal invariant quantum
field theory in two and more dimensions' in {\em Nonperturbative quantum field
theory: Proc. NATO Advanced Summer Institute (Carg\`ese, 1987)} ed. G. 't Hooft, A. Jaffe, G.
Mack, P.K. Mitter and R. Stora (Plenum Press, New York, 1988).

\bibitem{Milnor} J. Milnor, `Remarks on infinite-dimensional Lie
groups', in {\em Relativity, Groups and Topology II}, Les Houches
Session XL, 1983, ed. B.S. DeWitt and
R. Stora (North-Holland, Amsterdam, 1984). 

\bibitem{MorrisThorne} M.S. Morris and K.S. Thorne, `Wormholes in
spacetime and their use for interstellar travel: A tool for teaching
general relativity', Am. J. Phys., {\bf 56} (1988) 395--412.

\bibitem{Neeb}    K.-H. Neeb, 
                  `Central extensions of infinite-dimensional Lie groups',
                  Ann. Inst. Fourier, Grenoble, {\bf 52} (2002) 1365--1442.

\bibitem{OlumGraham} K.D. Olum and N. Graham, `Static negative energies near a domain wall'
Phys. Lett. B {\bf 554} (2003) 175--179.

\bibitem{Pfenning_em} M.J. Pfenning, `Quantum inequalities for the electromagnetic
field', Phys. Rev. D {\bf 65}, 024009 (2002).  

\bibitem{FPstat} M.J. Pfenning and L.H. Ford, `Scalar field quantum inequalities in static spacetimes',
                 Phys. Rev. D {\bf 57} (1998) 3489--3502.

\bibitem{FPwarp} M.J. Pfenning and L.H. Ford, `The unphysical nature of ``warp drive''', 
                 Class. Quantum Grav. {\bf 14} (1997) 1743--1751.

\bibitem{PS86}   A. Pressley and G. Segal, {\em Loop Groups}, 
                  (Oxford University Press, Oxford, 1999).

\bibitem{Puk}    L. Puk\'anszky, `The Plancherel formula for the universal covering group of
${\rm SL}(R,2)$',
                 Math. Ann., {\bf 156} (1964) 96--143.

\bibitem{RSi} M. Reed and B. Simon, {\it Methods of modern mathematical physics, Vol. 1:
functional analysis}, (Academic Press, New York, 1972).

\bibitem{RSii} M. Reed and B. Simon, {\it Methods of modern mathematical physics, Vol. 2:
Fourier analysis, self-adjointness}, (Academic Press, New York, 1975).

\bibitem{Roman_review} T.A. Roman, `Some thoughts on energy conditions and wormholes',
to appear in the Proceedings of 
the Tenth Marcel Grossmann Meeting on General Relativity and Gravitation {\tt gr-qc/0409090}.

\bibitem{Schott} M. Schottenloher, `A mathematical introduction to
conformal field theory', Lecture Notes in Physics m43, (Springer-Verlag,
Berlin, 1997). 

\bibitem{Segal81} G. Segal, `Unitary representations of some infinite
dimensional groups',
                  Commun. Math. Phys., {\bf 80} (1981) 301--342.

\bibitem{StrWigh} R.F. Streater and A.S. Wightman, 
                  {\em PCT, spin and statistics, and all that}, 
                  (Princeton University Press, Princeton, 1978)

\bibitem{TL}      V. Toledano Laredo, 
                  `Integrating unitary representations of infinite-dimensional Lie groups',
                  J. Funct. Anal. {\bf 161} (1999), 478--508.

\bibitem{Varadarajan} V.S. Varadarajan, {\em Geometry of Quantum Theory, Vol
II: Quantum Theory of Covariant Systems} (Van Nostrand, New York, 1970). 

\bibitem{VerchANEC} R. Verch, `The averaged null energy condition for
general quantum field theories in two dimensions',
J. Math. Phys., {\bf 41} (2000) 206--217.

\bibitem{Vollick} D.N. Vollick, `Quantum inequalities in curved two-dimensional spacetimes'
                  Phys. Rev. D{\bf 61} (2000) 084022.

\bibitem{Witten84} E. Witten, `Nonabelian bosonization in two dimensions', Commun. Math. Phys.,
{\bf 92}  (1984) 455--472.

\bibitem{YuWu}   H. Yu, P. Wu, `Quantum inequalities for the free Rarita-Schwinger fields in
flat spacetime', Phys. Rev. D {\bf 69} (2004) 064008.

\bibitem{Zuber} J.-B. Zuber, `CFT, BCFT, ADE and all that', {\tt hep-th/0006151}.


\end{thebibliography}
\end{document}